\documentclass[reprint,amssymb,amsmath,superscriptaddress,aps,showpacs,10pt,floatfix,longbibliography,prb]{revtex4-2}

\usepackage[pdftex]{graphicx} 
\usepackage{epstopdf}
\usepackage{verbatim}
\usepackage{amsmath}
\usepackage{color}
\usepackage{subfigure}
\usepackage{amsbsy}
\usepackage{wasysym}
\usepackage{textcomp}
\usepackage{times}
\usepackage{float}
\usepackage{latexsym,amsmath,amssymb,bm,euscript}
\usepackage[colorlinks=true,linkcolor=blue,citecolor=blue]{hyperref}
\usepackage{hyperref}
\usepackage{soul}
\usepackage[normalem]{ulem}
\usepackage{mathrsfs}
\usepackage{lettrine}
\usepackage{xspace}
\usepackage{filecontents}

\begin{document}

\title{Gapless Spin Liquid Behavior in A Kagome Heisenberg Antiferromagnet\\ with Randomly Distributed Hexagons of Alternate Bonds}

\author{Jiabin Liu}
\thanks{These authors contributed equally to this work}
\affiliation{Wuhan National High Magnetic Field Center and School of Physics, Huazhong University of Science and Technology, 430074 Wuhan, China}

\author{Long Yuan}
\thanks{These authors contributed equally to this work}
\affiliation{Wuhan National High Magnetic Field Center and School of Physics, Huazhong University of Science and Technology, 430074 Wuhan, China}

\author{Xuan Li}
\thanks{These authors contributed equally to this work}
\affiliation{Institute of Physics, Chinese Academy of Sciences, P.O. Box 603, Beijing 100190, China}
\affiliation{University of Chinese Academy of Sciences, Beijing 100049, China}

\author{Boqiang Li}
\affiliation{Wuhan National High Magnetic Field Center and School of Physics, Huazhong University of Science and Technology, 430074 Wuhan, China}

\author{Kan Zhao}
\affiliation{School of Physics, Beihang University, Beijing 100191, China}

\author{Haijun Liao}
\email{navyphysics@iphy.ac.cn}
\affiliation{Institute of Physics, Chinese Academy of Sciences, P.O. Box 603, Beijing 100190, China}
\affiliation{Songshan Lake Materials Laboratory, Dongguan, Guangdong 523808, China}

\author{Yuesheng Li}
\email{yuesheng\_li@hust.edu.cn}
\affiliation{Wuhan National High Magnetic Field Center and School of Physics, Huazhong University of Science and Technology, 430074 Wuhan, China}

\date{\today}

\begin{abstract}
We demonstrate that the new single crystal of YCu$_3$[OH(D)]$_{6.5}$Br$_{2.5}$ (YCOB) is a kagome Heisenberg antiferromagnet (KHA) without evident orphan spins ($\ll$ 0.8\%). The site mixing between polar OH$^-$ and non-polar Br$^-$ causes local distortions of Cu-O-Cu exchange paths, and gives rise to 70(2)\% of randomly distributed hexagons of alternate bonds ($\sim$ $J_1-\Delta J$ and $J_1+\Delta J$) and the rest of almost uniform hexagons ($\sim$ $J_1$) on the kagome lattice. Simulations of the random exchange model with $\Delta J$/$J_1$ = 0.7(1) show good agreement with the experimental observations, including the weak upturn seen in susceptibility and the slight polarization in magnetization. Despite the average antiferromagnetic coupling of $J_1$ $\sim$ 60 K, no conventional freezing is observed down to $T$ $\sim$ 0.001$J_1$, and the raw specific heat exhibits a nearly quadratic temperature dependence below 1 K $\sim$ 0.02$J_1$, phenomenologically consistent with a gapless (spin gap $\leq$ 0.025$J_1$) Dirac quantum spin liquid (QSL). Our result sheds new light on the theoretical understanding of the randomness-relevant gapless QSL behavior in YCOB, as well as in other relevant materials.
\end{abstract}

\maketitle

\section{Introduction}
The search for quantum spin liquid (QSL) has become one of the central topics in the fields spanning experiment and theory since Anderson's initial proposal~\cite{anderson1973resonating}. QSLs are intimately related to the realizing of the topological quantum computation~\cite{nayak2008non} and the understanding of the high temperature superconductivity~\cite{anderson1987resonating} due to exotic fractional excitations~\cite{kasahara2018majorana} and other emergent properties~\cite{balents2010spin,Broholmeaay0668,RevModPhys.89.025003,Li2019YbMgGaO4}. The $S$ = $\frac{1}{2}$ kagome Heisenberg antiferromagnet (KHA)~\cite{PhysRevLett.62.2405,Waldtmann1998First,PhysRevLett.98.117205,PhysRevLett.118.137202,PhysRevX.7.031020,2018Entanglement,PhysRevB.98.224414,Yan2011,PhysRevLett.109.067201} is one of the most promising candidates hosting QSLs. Despite recent progress, the precise nature of the QSL in the KHA remains elusive. Previous density-matrix renormalization group (DMRG) studies imply a fully gapped $\mathbb{Z}_2$ QSL ground state (GS)~\cite{Yan2011,PhysRevLett.109.067201}, but variational Monte Carlo~\cite{PhysRevLett.98.117205,PhysRevB.87.060405,PhysRevB.89.020407}, tensor network~\cite{PhysRevLett.118.137202}, and more recent DMRG~\cite{PhysRevX.7.031020} calculations suggest a gapless Dirac QSL.

Experimentally, many prominent $S$ = $\frac{1}{2}$ KHAs have been extensively studied, including ZnCu$_3$(OH)$_6$Cl$_2$ (herbertsmithite)~\cite{Shores2010A,PhysRevLett.98.107204,PhysRevB.76.132411,PhysRevLett.100.157205,han2012fractionalized,PhysRevB.94.060409,PhysRevLett.100.087202,Fu2015Evidence,Khuntia2020Gapless}, $\alpha$-Cu$_3$Zn(OH)$_6$Cl$_2$ (kapellasite)~\cite{PhysRevLett.109.037208,PhysRevB.90.205103}, Cu$_3$Zn(OH)$_6$FBr~\cite{feng2017gapped,fu2018dynamic}, [NH$_4$]$_2$[C$_7$H$_{14}$N][V$_7$O$_6$F$_{18}$]~\cite{Aidoudi2011an,PhysRevLett.110.207208}, (CH$_3$NH$_3$)$_2$NaTi$_3$F$_{12}$~\cite{Jiang2020Synthesis}, ZnCu$_3$(OH)$_6$SO$_4$~\cite{li2014Gapless,PhysRevLett.119.137205,gom2019Kondo}, etc. They show no conventional magnetic transitions down to the lowest measuring temperature, despite the strong exchange couplings. However, most of these candidates suffer from 4--27\% of magnetic defects~\cite{Khuntia2020Gapless,PhysRevB.90.205103,PhysRevB.94.060409,feng2017gapped,PhysRevLett.110.207208,Jiang2020Synthesis,li2014Gapless}. Orphan spins created by these magnetic defects contribute a pronounced Curie-like tail of magnetic susceptibility~\cite{PhysRevB.76.132411,PhysRevLett.110.207208} and Schottky-like anomaly of specific heat~\cite{PhysRevLett.100.157205,PhysRevLett.110.207208} at low temperatures ($\leq$ 100 K), and thus prevent us from probing the intrinsic low-energy properties ~\cite{RevModPhys.88.041002}. Therefore, achieving ultrahigh-quality materials is still the key challenge for the study of kagome QSLs~\cite{Broholmeaay0668}.

Recently, a new kagome QSL candidate, nondeuterated YCu$_3$(OH)$_{6.5}$Br$_{2.5}$, had been successfully synthesized~\cite{CHEN2020167066} by Chen \emph{et al.} The magnetic Cu$^{2+}$ ($S$ = $\frac{1}{2}$) ions were expected to fully occupy the regular kagome sites, and be free from the site mixing with other nonmagnetic ions~\cite{CHEN2020167066}. Despite the strong antiferromagnetic coupling, YCu$_3$[OH(D)]$_{6.5}$Br$_{2.5}$ (YCOB) exhibits no magnetic transition down to 50 mK~\cite{CHEN2020167066,arXiv:2107.11942}. However, the low-energy magnetism of this promising QSL candidate YCOB remains poorly understood.

In this paper, we report a comprehensive study of the frustrated magnetism on the high-quality single crystal of YCOB, including sub-kelvin thermodynamic \& electron spin resonance (ESR) measurements and microscopic modeling by density functional theory (DFT) and state-of-the-art quantum many-body computations. The nonsymmetric distribution of OH$^-$/Br$^-$ pushes 70(2)\% of Y$^{3+}$ away from its ideal position, and gives rise to two alternate local distortions of the Cu-O-Cu superexchange paths ($\sim$ $J_1-\Delta J$ and $J_1+\Delta J$) around the hexagons on the kagome lattice, whereas the rest of symmetric local environments result in nearly uniform hexagons ($\sim$ $J_1$). Through fitting the magnetic susceptibilities, we find a profound fluctuation of the exchange couplings, $\Delta J$/$J_1$ $\sim$ 0.7. The weak upturn in the low-$T$ susceptibility and the slight polarization in magnetization originate from the local moments induced by the bond randomness. Despite the average antiferromagnetic interaction of $J_1$ $\sim$ 60 K, no magnetic freezing is found down to 50 mK~\cite{arXiv:2107.11942} $\sim$ 0.001$J_1$. Below $\sim$ 0.02$J_1$, the specific heat exhibits a nearly quadratic temperature dependence, which is phenomenologically consistent with the predictions of a gapless U(1) Dirac QSL or a $\mathbb{Z}_2$ QSL with a gap less than 0.025$J_1$. We argue that the bond randomness is a critical ingredient for the gapless QSL behavior observed in YCOB, as well as in other relevant materials.

\begin{figure}
\includegraphics[width=8.5cm,angle=0]{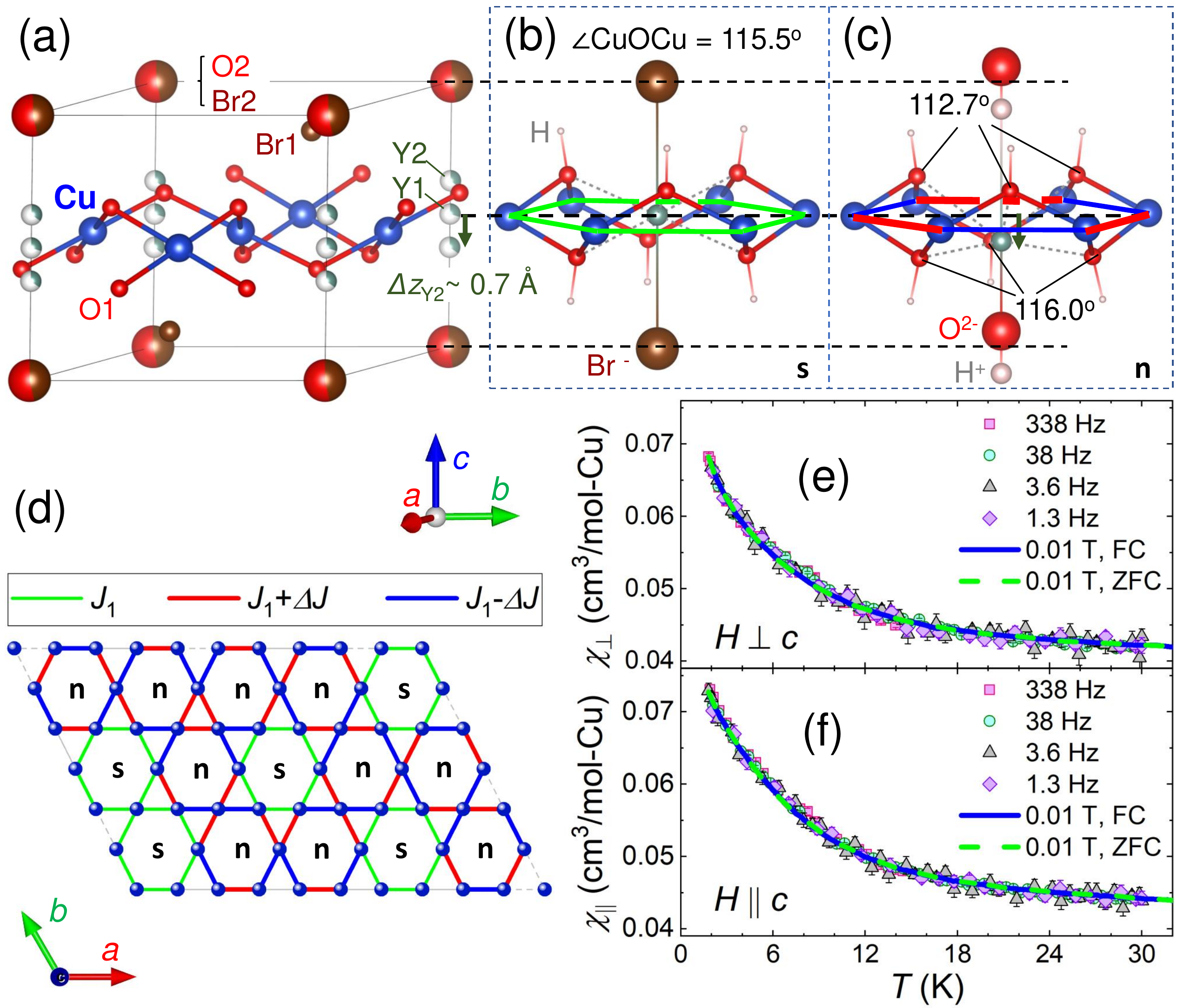}
\caption{(Color online)
(a) The experimental lattice structure of YCOB. (b, c) Optimized crystal structures for supercells containing the symmetric (s) and nonsymmetric (n) environments, respectively. The nonsymmetric environments (probability $\sim$ 70\%) cause the deviations of Y$^{3+}$ from its ideal positions (green arrows) and the alternate superexchanges (alternate $\angle$CuOCu are marked in c) along the hexagons on the kagome lattice of Cu$^{2+}$. (d) Sketch of the model no.2 with $\sim$ 30\% of uniform hexagons (green) and $\sim$ 70\% of randomly distributed hexagons of alternate bonds (blue and red). Real part of the ac susceptibilities measured on YCOB along the $ab$ plane (e) and $c$ axis (f), respectively. The dashed and solid lines present the dc susceptibilities measured at 0.01 T under zero-field cooling (ZFC) and field cooling (FC), respectively.}
\label{fig1}
\end{figure}

\section{Experimental Details}
We performed a recrystallization in a temperature gradient of $\sim$ 2 $^{\circ}$C/cm for a month, and obtained transparent green single crystals of YCOB (typical size 3$\times$3$\times$0.3~mm) (see Appendix~\ref{sec1}). Magnetization up to 14 T and dc/ac susceptibilities down to 1.8 K were measured in a physical property measurement system (PPMS) and a magnetic property measurement system, respectively.  The specific heat down to 0.36 K~\footnote{The Schottky anomaly of the H and D nuclear spins can appear in $C_\mathrm{p}$ below $\sim$ 0.2 K~\cite{arXiv:2107.11942,Yamashita2011Gapless,li2014Gapless}, and the temperature of 0.36 K $\sim$ 0.006$J_1$ is low enough for YCOB that the entropy flattens out at $S_\mathrm{m}$ $<$ 1.3\%$R$ln 2 [see Fig.~\ref{figs12}(d)].} was measured in a PPMS using a $^{3}$He refrigerator. The ESR spectra were measured on continuous wave spectrometers at x-band frequencies ($\sim$ 9.7 GHz). 

\begin{figure}
\begin{center}
\includegraphics[width=8.6cm,angle=0]{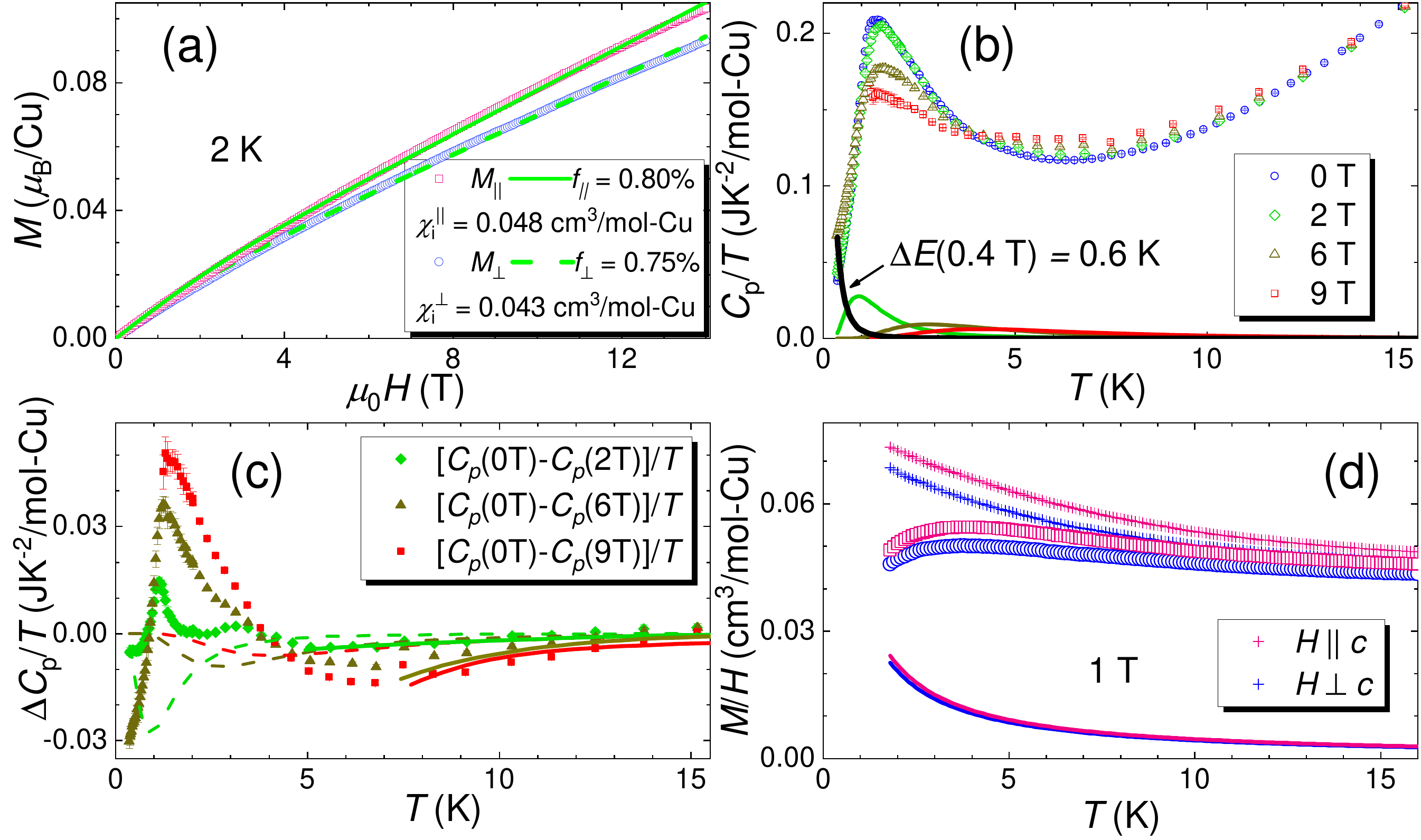}
\caption{(a) Magnetization measured at 2 K on the YCOB single crystal with the magnetic fields applied parallel and perpendicular to the $c$ axis. The green lines present the Brillouin fits [see Eq.~(\ref{eq_free1})]. (b) Raw specific heat of YCOB measured in selected magnetic fields applied along the $c$ axis, with the lines showing the calculated Schottky functions for 0.8\% of free spins [see Eq.~(\ref{eqS6})]. (c) The difference between the zero- and nonzero-field specific heat. The dashed lines show the calculations of 0.8\% of free spins, whereas the solid lines present the 21-site FLD calculations of the random model no.2 that is valid above $\sim$ 0.1$J_1$ $\sim$ 7 K. (d) Magnetic susceptibilities of YCOB measured at 1 T applied along the $ab$ plane and $c$ axis (cross). The hollow scatters show the data obtained by subtracting the Brillouin functions for 0.75\% and 0.80\% of ``free spins'' (colored lines) from the corresponding raw susceptibilities.}
\label{fig_free}
\end{center}
\end{figure}

The DFT$+U$ calculations were performed in the Vienna Ab initio Simulation Package (VASP)~\cite{vasp1,vasp2} (see Appendix~\ref{sec2}). We carried out the finite-temperature Lanczos diagonalization (FLD) simulations for the thermodynamic properties of YCOB, as well as the local susceptibility $\chi_i^{\mathrm{loc}}$ and correlation function. The FLD calculations were conducted on the 18-, 21-, 24-, and 27- site KHA clusters with periodic boundary conditions (PBC), we excluded the thermodynamic data calculated below $T_{\mathrm{min}}$ in the main text, where the finite-size effect is significant (see Appendix~\ref{sec3}). We fit the experimental data by minimizing the residual function,
\begin{equation}
R_p=\sqrt{\frac{1}{N_0}\sum_{i}(\frac{X_i^{\mathrm{obs}}-X_i^{\mathrm{cal}}}{\sigma_i^{\mathrm{obs}}})^2},
\label{eq0}
\end{equation}
where $N_0$, $X_i^{\mathrm{obs}}$ and $\sigma_i^{\mathrm{obs}}$ are the number of the data points, the observed value and its standard deviation, respectively, whereas $X_i^{\mathrm{cal}}$ is the calculated value. The international system of units is used throughout this work.

\section{Results and Discussion}
\subsection{Exchange Hamiltonian}
Fig.~\ref{fig1}(a) shows the crystal structure of YCOB, where the magnetic Cu$^{2+}$ ($S$ = $\frac{1}{2}$) ions occupy the 3$f$ Wyckoff position of the space group $P$\={3}$m$1, and form the kagome lattice [see Fig.~\ref{fig1}(d)]. The site mixing between the magnetic Cu$^{2+}$ (ionic radius $r_{\mathrm{Cu}^{2+}}$ = 0.73 {\AA}) and other nonmagnetic ions, i.e., Y$^{3+}$ ($r_{\mathrm{Y}^{3+}}$ = 0.90 {\AA}), H(D)$^{+}$, O$^{2-}$, Br$^{-}$, is unlikely due to the large chemical differences, which has been confirmed by single-crystal x-ray diffraction (XRD)~\cite{CHEN2020167066,arXiv:2107.11942}.

\begin{figure}
\includegraphics[width=8.2cm,angle=0]{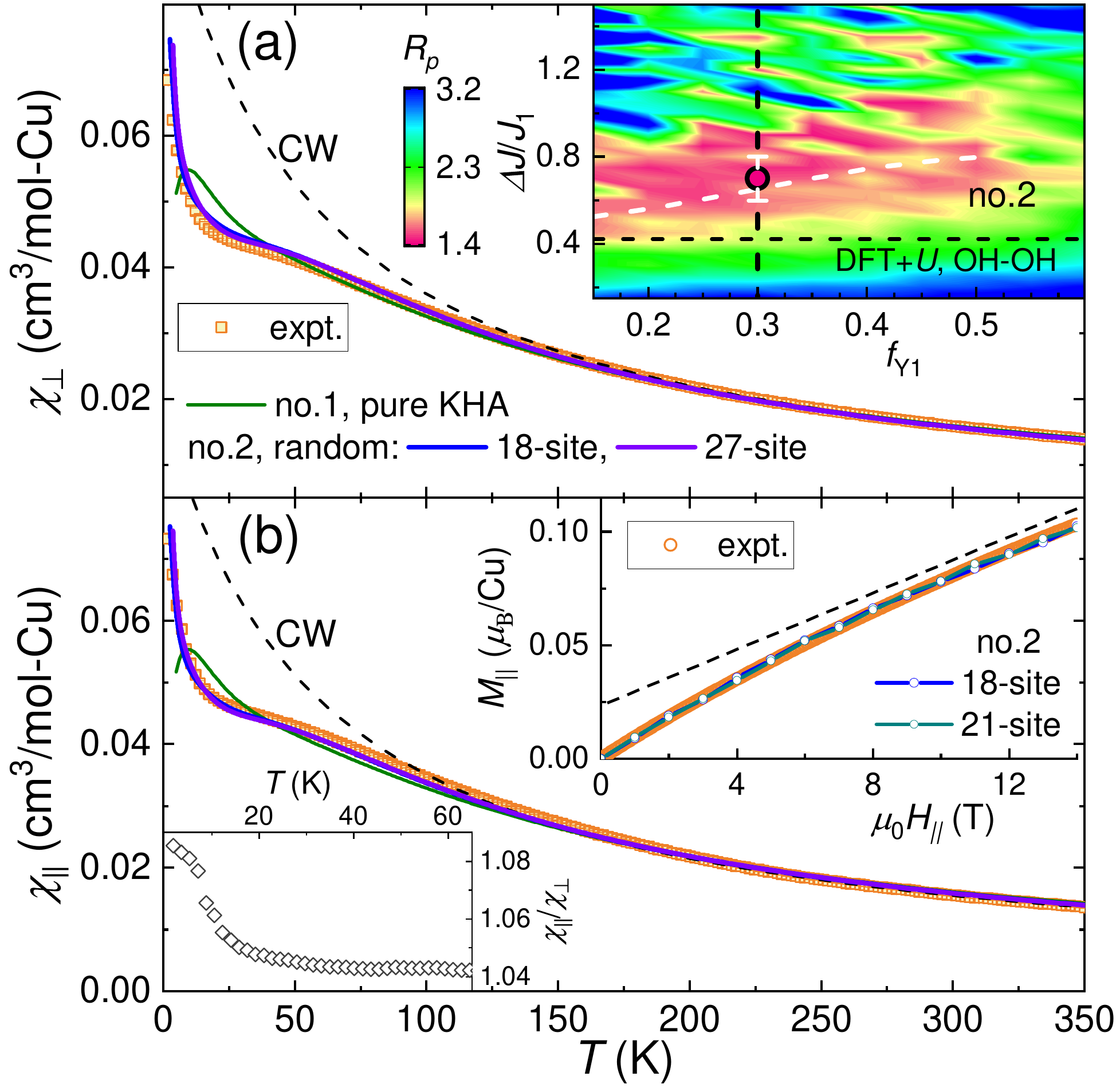}
\caption{(Color online)
Combined fits to the raw susceptibilities measured on YCOB along the $ab$ plane (a) and $c$ axis (b), respectively, using models nos.1 (ideal KHA) and 2 (with randomly distributed hexagons of alternate bonds). The dashed lines show the Curie-Weiss fits above 150 K. The inset of (a) presents the deviation $R_p$ of using model no.2, with the dotted white line showing the optimized parameters. The vertical dashed line presents the occupancy fraction of Y1 from XRD, $f_{\mathrm{Y}1}$ = 0.30(2), and the horizontal line displays the bond randomness ($\Delta J$/$J_1$) calculated by DFT$+U$ in the OH-OH configuration [Fig.~\ref{fig1} (c)]. The circle shows the optimized value, $\Delta J$/$J_1$ = 0.7(1). The upper inset of (b) presents the magnetization measured at 2 K, with showing the 18- and 21-site FLD calculations of model no.2. The dashed straight line marks a linear dependence. The lower inset of (b) displays the measured magnetic anisotropy ($\chi_{\parallel}$/$\chi_{\perp}$).}
\label{fig2}
\end{figure}

From the Curie-Weiss fitting, we get the Curie-Weiss temperatures $\theta_{\perp}$ = $-$57.5(4) K and $\theta_{\parallel}$ = $-$58.7(3) K (see Figs.~\ref{fig2} and~\ref{figs2}), which indicate a strong antiferromagnetic exchange coupling, $J_1$ $\sim$ 60 K~\cite{CHEN2020167066}. A weak upturn is found in the susceptibilities below $\sim$ 30 K $\sim$ 0.5$J_1$ [see Figs.~\ref{fig1}(e) and \ref{fig1}(f)], which speaks against the simplest nearest-neighbor (NN) KHA model (see Fig.~\ref{fig2})~\footnote{Because the ideal NN KHA exhibits a broad peak in susceptibility at $T_\mathrm{p}$ $\sim$ 0.154$J_1$ $\sim$ 9 K}, and suggests the existence of net magnetic moments. Unlike many other kagome QSL candidates~\cite{PhysRevB.76.132411,li2014Gapless}, e.g., ZnCu$_3$(OH)$_6$Cl$_2$, the model of conventional orphan spins fails to explain all the low-$T$ thermodynamic observations (see Fig.~\ref{fig_free}), which suggests the concentration of orphan spins is much lower than 0.8\% in YCOB (see Appendix~\ref{sec3}). Therefore, the low-$T$ behaviors are governed by the intrinsic properties of the specific KHA of YCOB, which deviates from the ideal NN KHA.

To understand the frustrated magnetism of the KHA YCOB, we consider the following spin-1/2 Hamiltonian in an applied magnetic field of $\mathbf{H}$,
\begin{equation}
\mathcal{H}=\sum_{\langle ij\rangle}J_{1,ij}\mathbf{S}_i\cdot\mathbf{S}_j-\mu_0\mu_\mathrm{B}\mathbf{H}\cdot\mathbf{g}\cdot\sum_{i}\mathbf{S}_i+\mathcal{H'},
\label{eq1}
\end{equation}
where $\mathbf{g}$ is the diagonal matrix of $g$ factor with $g_{\perp}$ = 2.18(1) and $g_{\parallel}$ = 2.19(1) determined by fitting the magnetic susceptibility data (see Fig.~\ref{fig2}), and $\mathcal{H'}$ represents symmetry-allowed perturbations, including the Dzyaloshinsky-Moriya (DM) interaction~\cite{PhysRevLett.101.026405,PhysRevLett.125.027203}, further-neighbor couplings~\cite{li2021continuous}, XXZ anisotropy~\cite{PhysRevLett.108.157202}, etc.

Starting with the simplest NN KHA (see Fig.~\ref{fig2} for model no.1), we fit the magnetic susceptibilities ($\chi_{\perp}$ and $\chi_{\parallel}$) measured on YCOB above $\sim$ 0.1$J_1$ $\sim$ 7 K, and find $J_1$ = 61.6 K with the least $R_p$ $\sim$ 3.3~\cite{PhysRevB.98.094423}. However, the weak upturn of $\chi_{\perp}$ and $\chi_{\parallel}$ below $\sim$ 30 K is poorly understood (see Fig.~\ref{fig2}).

The measured magnetic anisotropy, $\chi_{\parallel}$/$\chi_{\perp}$, only increases by $\sim$ 4\% from $T$ $\sim$ $J_1$ down to 1.8 K $\sim$ 0.03$J_1$ [see the lower inset of Fig.~\ref{fig2}(b)], suggesting that the interaction anisotropy isn't critical in YCOB due to the weak spin-orbit coupling of the 3$d$ electrons. Based on the DFT+$U$ calculation, we find that both further-neighbor and interlayer exchange couplings are negligible, which are less than 4\% of $J_1$ (highly similar to the sibling YCu$_3$(OH)$_6$Cl$_3$~\cite{PhysRevLett.125.027203}). Therefore, we mainly restrict ourselves to the effective models with only NN Heisenberg couplings at this stage.

In YCOB, the site mixing between OH$^-$ and Br$^-$ occurs around the ideal Y$^{3+}$ site (Wyckoff position 1$b$ at the center of each hexagon on the kagome lattice of Cu$^{2+}$), based on the XRD~\cite{CHEN2020167066,arXiv:2107.11942}. Although OH$^-$ and Br$^-$ have the same net charge, OH$^-$ is polar, and causes local distortions that may have a strong influence on the exchange couplings. To go beyond the average structure from XRD, we performed the DFT calculations by constructing various OH/Br configurations (with H) to simulate this site mixing. In the symmetric (e.g., Br-Br) configurations, the center of inversion is located at 1$b$ [see Fig.~\ref{fig1}(b)], and Y$^{3+}$ occupies its ideal position with a statistical probability $f_{\mathrm{Y}1}$ = 30(2)\% from XRD (see Table~\ref{table1})~\cite{CHEN2020167066}. In contrast, the other nonsymmetric OH/Br configurations, e.g., OH-OH, push Y$^{3+}$ away from its ideal position, $\Delta z_{\mathrm{Y}2}$ $\sim$ 0.1$c$ $\sim$ 0.6{\AA}, well consistent with the XRD result [see Figs.~\ref{fig1}(a) and~\ref{fig1}(c)]. More importantly, the nonsymmetric configurations give rise to the alternate bond angles $\angle$CuOCu along the hexagons of the kagome lattice [see Fig.~\ref{fig1}(c)]. Therefore, $1-f_{\mathrm{Y}1}$ = 70(2)\% of randomly distributed hexagons with alternate bonds ($\sim$ $J_1-\Delta J$ and $J_1+\Delta J$) should be expected on the kagome lattice of YCOB [see Fig.~\ref{fig1}(d)]~\cite{PhysRevB.57.5326}.

\begin{figure*}
\includegraphics[width=16cm,angle=0]{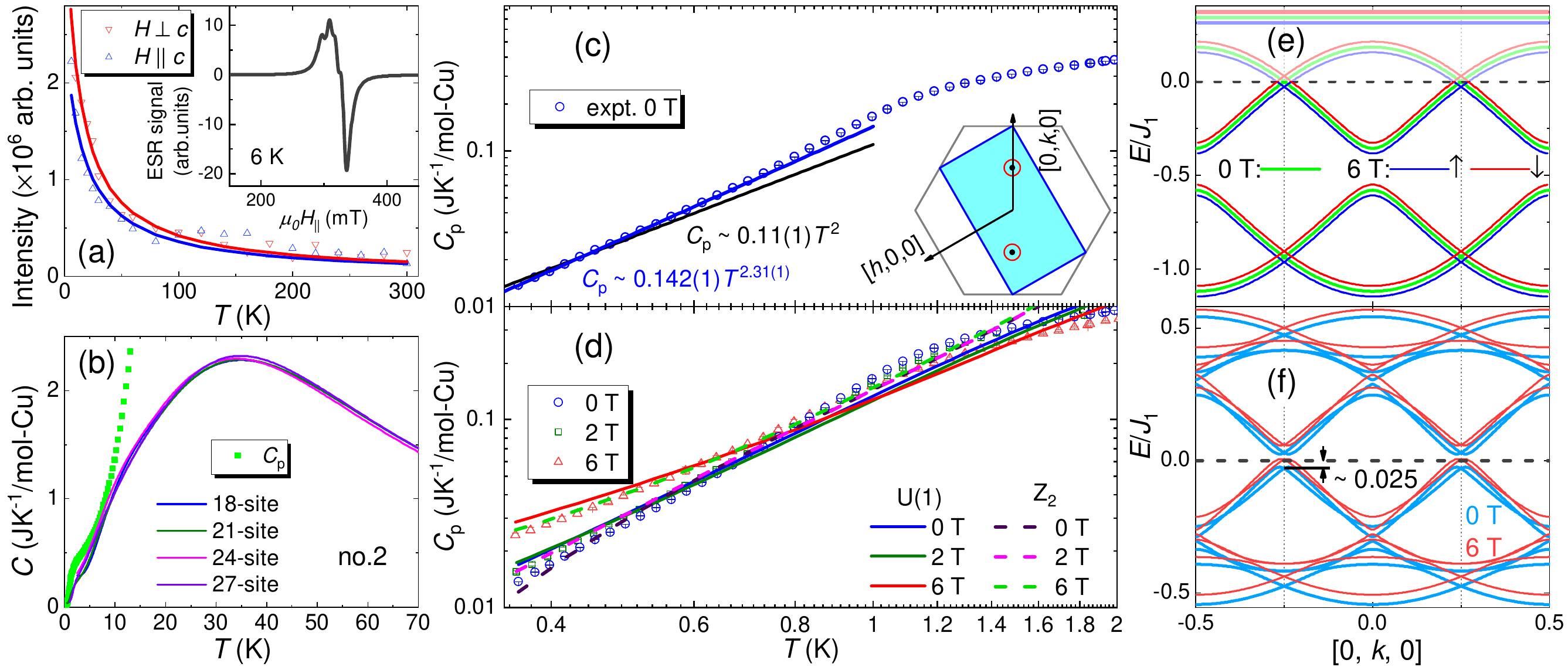}
\caption{(Color online)
(a) ESR intensities of YCOB. The color lines present the Curie-Weiss fits, and the inset shows the ESR spectrum measured at 6 K along the $c$ axis. (b) Magnetic specific heat calculated by using model no.2 ($\Delta J$/$J_1$ = 0.7) on various size KHA clusters, with the experimental total specific heat for comparison. (c) Raw specific heat of YCOB measured at 0 T. The blue and black lines present the power-law and quadratic behaviors, respectively. The inset shows the Brillouin zone of the mean-field Ansatz on the kagome lattice, and the circles mark the positions of the Dirac nodes. (d) Raw specific heat measured in selected magnetic fields applied along the $c$ axis. The thin and thick lines display the fittings below 1 K, using the U(1) and $\mathbb{Z}_2$ mean-field Ansatz, respectively. (e) Band structure of the U(1) Dirac QSL at $h$ = 0. The field of 6 T splits the zero-field bands (green) into the spin-up ($\uparrow$, blue) and spin-down ($\downarrow$, red) bands. (f) Band structure of the $\mathbb{Z}_2[0,\pi]\beta$ QSL in the Nambu representation at $h$ = 0. The zero-field spin gap is marked.}
\label{fig3}
\end{figure*}

We further conduct the DFT$+U$ calculation to estimate the exchange couplings for YCOB, and find that the fluctuation of these alternate bonds, $\Delta J$/$J_1$, is strong, ranging from 0 to $\sim$ 1.3 (see Appendix~\ref{sec2}). According to the simplified model of the crystal structure [see Fig.~\ref{fig1}(a)], we construct a simple KHA model (no.2) with $1-f_{\mathrm{Y}1}$ = 70\% of randomly distributed hexagons of alternate bonds and the rest of almost uniform hexagons on the kagome lattice, as illustrated in Fig.~\ref{fig1}(d). The real situation of YCOB may be more complicated, but one always seeks to explain the bulk of experimental observations within the minimum model that captures the essential physics.

Using the model no.2, we are able to fit the experimental susceptibilities very well at $J_1$ = 50(1) K and $\Delta J$/$J_1$ = 0.7(1) with a significantly reduced $R_p$ = 1.4(1) (see Fig.~\ref{fig2}). The finite-size effect is trivial in the model no.2 [see Figs.~\ref{fig2} and \ref{fig3}(b)], and both the weak upturn below $\sim$ 0.5$J_1$ and the broad hump at $\sim$ $J_1$ can be excellently reproduced. In contrast, the complete and continuous randomness of $J_1$ and $g$ with a Lorentzian distribution~\cite{PhysRevX.10.011007,Li2021spin} fails to fit the experimental susceptibilities (see Fig.~\ref{figs7}). Besides that, the model no.2 also well simulates the slight polarization seen in the low-$T$ magnetization without any parameter tuning [the upper inset of Fig.~\ref{fig2}(b)]. These observations strongly support the formation of effective $\sim$ 0.8\% (see Fig.~\ref{fig_free}) of local moments due to the bond randomness (see Fig.~\ref{fig4})~\cite{2014Quantum,PhysRevB.77.184423,PhysRevB.79.214415,PhysRevLett.104.177203}. The temperature dependence of the ESR intensity shows a Curie-Weiss behavior with the Curie-Weiss temperatures, $\theta_{\perp}^{\mathrm{ESR}}$ = $-$12(2) K and $\theta_{\parallel}^{\mathrm{ESR}}$ = $-$16(4) K [Figs.~\ref{fig3}(a) and \ref{figs4}(d)], suggesting the ESR lines may originate from the nearly free local moments induced by the bond randomness~\cite{PhysRevB.92.134407}. Moreover, the ESR spectra of YCOB are so narrow that the hyperfine structures are clearly visible (see Fig.~\ref{figs3}), implying that the weak coupling between the randomness-induced moments may be nearly isotropic Heisenberg~\cite{PhysRevLett.101.026405}.    

It is worth to note that the nearly free local moments induced by bond randomness in YCOB are from the nontrivial many-body correlations and random-singlet physics (see Fig.~\ref{fig4})~\cite{PhysRevLett.104.177203,2014Quantum,2014Quantumspin,PhysRevB.92.134407}, which is essentially different from the trivial “orphan” spins observed in other existing kagome QSL candidate materials. Above 0.02$J_1$ $\sim$ 1 K, the theory of the scaling collapse recently proposed by Kimchi et al.~\cite{Kimchi2018Scaling} is applicable to YCOB, confirming the formation of random-singlet state. However, below $\sim$ 1 K the raw zero-field specific heat shows a power-law behavior $C_\mathrm{p}$ $\propto$ $T^{\alpha}$ with $\alpha$ = 2.31(1) $>$ 1 [Fig.~\ref{fig3}(c)] and the existing theories obviously fail (see Fig.~\ref{figs13}).

\subsection{Sub-Kelvin specific heat and mean-field Ansatz}
Below $\sim$ 2 K, the lattice contribution is completely negligible [see Figs.~\ref{figs11}(a) and \ref{figs11}(b)] in the Mott insulator YCOB with room-$T$ resistance larger than 20 M$\Omega$ parallel and perpendicular to the $c$ axis, and thus the raw specific heat directly measures the intrinsic magnetic properties that are sensitive to the low-energy ($\leq$ 3\%$J_1$) density of states.

Despite the strong antiferromagnetic exchange coupling of $J_1$ $\sim$ 60 K, no conventional magnetic freezing is evidenced, as there is neither splitting of zero-field-cooling and field-cooling dc susceptibilities nor frequency dependence of the ac susceptibilities down to 1.8 K ($\sim$ 0.03$J_1$) [see Figs.~\ref{fig1}(e) and \ref{fig1}(f)]~\cite{CHEN2020167066}. No sharp ``$\lambda$'' specific heat peak was observed down to 0.36 K ($\sim$ 0.006$J_1$) [see Fig.~\ref{fig3}(c)]. Moreover, our recent thermal conductivity measurements show no sign of phase transitions down to $\sim$ 80 mK~\cite{thermalCon}. These observations consistently show that YCOB is a very promising (randomness-induced) QSL candidate without evident orphan spins, and thus one is able to explore in depth its intrinsic GS nature. Thereby, we analysis the experimental results by using the three most promising mean-field QSLs classified by the Projective Symmetry Group method~\cite{PhysRevB.65.165113,PhysRevLett.98.117205,PhysRevB.83.224413}, including the uniform RVB~\cite{PhysRevLett.98.117205}, U(1) Dirac~\cite{PhysRevLett.98.117205}, and $\mathbb{Z}_2[0,\pi]\beta$~\cite{PhysRevB.83.224413} states. 

\begin{figure}
\includegraphics[width=8.6cm,angle=0]{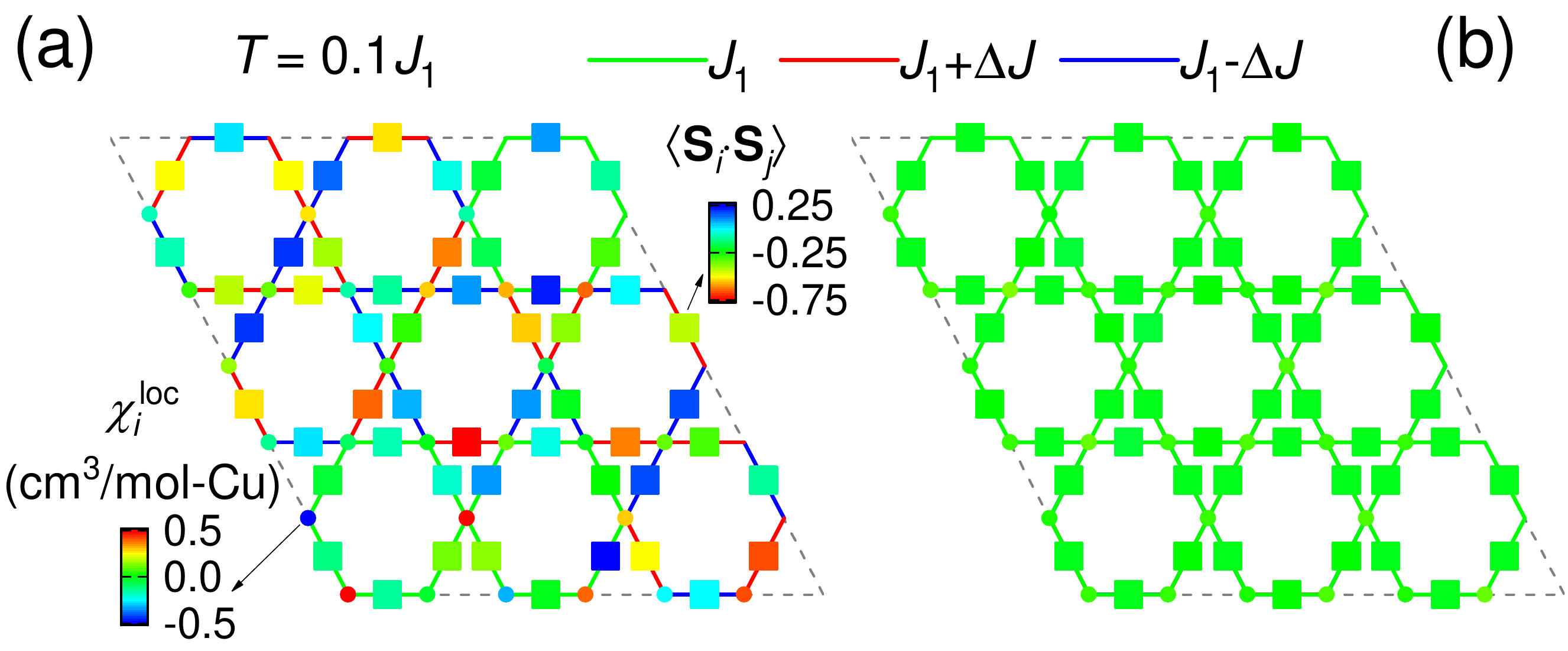}
\caption{(Color online)
Local susceptibility at each kagome site (see Ref.~\cite{PhysRevB.77.184423} for the definition) and correlation functions calculated for the 27-site samples of model no.2 with bond randomness (a) and the ideal KHA model no.1 (b), at $T$ = 0.1$J_1$~\footnote{$T$ = 0.1$J_1$ is low enough, as the correlation functions of both systems are nearly constant.} and $\mu_0H$ = 0 T. The dashed lines display the cluster with PBC.}
\label{fig4}
\end{figure}

The inherent exchange randomness should influence the GS nature of the spin system of YCOB. In contrast to the ideal KHA, the simulated GS of the random model indeed exhibits the signature of random-singlet or valence bond glass phase~\cite{PhysRevB.79.214415,PhysRevLett.104.177203}. The pair of NN spins with the strong exchange interaction ($J_1+\Delta J$) tends to form a singlet with extremely strong local correlation $\langle\bm{S}_i\cdot\bm{S}_j\rangle$ $\sim$ $-0.7$ (see Fig.~\ref{fig4}). However, the fraction of such singlets is still very low due to the strong frustration of the kagome lattice~\cite{PhysRevB.92.134407}, even in the presence of strong exchange randomness [see Fig.~\ref{fig4}(a)]. Therefore, the picture of mobile spinons is still applicable~\cite{PhysRevLett.104.177203}, and the mean-field Ansatz can nevertheless serve as a simple and phenomenological model for understanding the complicated GS nature of YCOB, despite the possible distribution of hopping strength (not as a direct function of the detailed distribution of the exchange couplings, see Fig.~\ref{fig4}). The distribution of hopping strength due to bond randomness might be general in almost all of the well-studied spin-liquid candidate compounds, $\kappa$-(ET)$_2$Cu$_2$(CN)$_3$, EtMe$_3$Sb[Pd(dmit)$_2$]$_2$, ZnCu$_3$(OH)$_6$Cl$_2$, etc.~\cite{PhysRevB.92.134407}.

The uniform RVB state gives rise to a linear temperature dependence of specific heat that obviously contradicts the observations in YCOB [see Fig.~\ref{figs14}(d)]. The zero-field specific heat of YCOB shows a nearly quadratic behavior [see Fig.~\ref{fig3}(c)], $C_\mathrm{p}\sim\gamma_2T^2$ with $\gamma_2$ = 0.11(1) JK$^{-3}$/mol-Cu, which is consistent with the U(1) Dirac state~\cite{PhysRevLett.98.117205}, with the NN hopping parameter $\chi_1$ = $\frac{k_\mathrm{B}}{J_1}\sqrt{\frac{6\sqrt{3}\zeta(3)R}{\pi\gamma_2}}$ = 0.29(1) and the Fermi velocity $v_\mathrm{F}$ = $\frac{aJ_1\chi_1}{\sqrt{2}\hbar}$ = 1.07(5)$\times$10$^3$ m/s~\cite{PhysRevLett.98.117205}. The resulted $\chi_1$ is slightly larger than the self-consistent mean-field value 0.221~\cite{PhysRevB.63.014413}. The mean-field theory fails to strictly impose the single-occupancy constraint~\cite{PhysRevB.72.045105,RevModPhys.89.025003}, and the unphysical states may result in the overestimated coefficient $\gamma_2$(MF) = 0.188 JK$^{-3}$/mol-Cu. This failure gets more severer when a magnetic field is applied (see below).

For the U(1) Dirac state, the applied magnetic field tends to form Fermi pockets [see Fig.~\ref{fig3}(e)], increase the low-energy density of states and thus low-$T$ specific heat, which qualitatively accounts for the observations [see Fig.~\ref{fig3}(d)]. However, the fermionic spinon with the effective magnetic moment $\mu_\mathrm{eff}$ $\sim$ $\mu_\mathrm{B}$ causes a much stronger dependence on the field in the mean-field theory, and can't fit well the specific heat measured in nonzero fields below $\sim$ 1 K [see Fig.~\ref{figs14}(h)]. A remedy is to set $\mu_\mathrm{eff}$ as an adjustable parameter, and we get $\mu_\mathrm{eff}$ $\sim$ 0.43$\mu_\mathrm{B}$ [see Fig.~\ref{fig3}(d)]~\footnote{The significant decrease of the effective magnetic moment from $\mu_\mathrm{B}$ may originate from the unphysical states due to the rough treatment of the single-occupied condition, which is another issue beyond the mean-field theory worth of further research.}. Moreover, the crossings of different specific heat curves measured in different magnetic fields are also well captured by this U(1) Dirac scenario [Fig.~\ref{fig3}(d)]. The real material of YCOB with bond randomness should have a distribution of hopping strength on the kagome lattice [see Fig.~\ref{fig4}(a)], which might slightly renormalize the band structure and the dispersion around the Dirac nodes and account for the observation of $\alpha$ = 2.31(1) $\neq$ 2.

In the neighborhood of the U(1) Dirac state, the $\mathbb{Z}_2[0,\pi]\beta$ QSL was proposed as the GS of the KHA as well~\cite{Yan2011,PhysRevLett.109.067201,PhysRevB.83.224413}. Therefore, we also fit the specific heat of YCOB by using the mean-field Ansatz of the $\mathbb{Z}_2[0,\pi]\beta$ state~\cite{PhysRevB.83.224413}, and obtain a similar band structure with a small zero-field spin (triplet) gap of $\sim$ 0.025$J_1$ [see Fig.~\ref{fig3}(f)] that is smaller than the DMRG value $\sim$ 0.05$J_1$~\cite{Yan2011}. The applied magnetic field tends to close the gap, form Fermi pockets, and thus increase the low-$T$ specific heat [see Fig.~\ref{fig3}(d)]. Similarly, by fitting we obtain the reduced $\mu_\mathrm{eff}$ $\sim$ 0.48$\mu_\mathrm{B}$. Finally, it is worth to note that the spin gap should be slightly overestimated, because the calculated data drops faster than the experimental one at $\sim$ 0.4 K as $T$ $\rightarrow$ 0 K at 0 T [see Fig.~\ref{fig3}(d)].

\subsection{Discussion}
In the sibling YCu$_3$(OH)$_6$Cl$_3$, no obvious antisite mixing of OH/Cl has been reported, the center of inversion is located at 1$b$, Y$^{3+}$ almost fully occupies its ideal position, i.e., $f_{\mathrm{Y}1}$ $\sim$ 1~\cite{Nokhrin2016Perfect}, and a long-range magnetic transition occurs at $T_\mathrm{N}$ = 12 K $\sim$ 0.15$J_1$~\cite{PhysRevMaterials.3.074401,PhysRevB.99.214441,PhysRevB.100.144420} possibly due to the DM interaction~\cite{PhysRevLett.125.027203}. In contrast, a gapless QSL behavior is observed in YCOB (similar to Ref.~\cite{arXiv:2107.11942}) with profound antisite mixing and thus bond randomness. Therefore, we believe that the bond randomness is a critical ingredient for stabilizing the quantum paramagnetic state. Both the susceptibility and specific heat of model no.2 indeed exhibit no sign of magnetic transition [see Figs.~\ref{fig2} and \ref{fig3}(b)], and a randomness-relevant gapless QSL state has been theoretically proposed in KHAs with a broad range of bond randomness~\cite{2014Quantum}. Since QSLs are typically fragile in the ideal cases~\cite{PhysRevLett.118.137202,PhysRevB.98.224414}, the bond randomness caused by site mixing may make the gapless QSL behavior more robust against various inevitable perturbations in real materials, e.g., ZnCu$_3$(OH)$_6$Cl$_2$ and YbMgGaO$_4$~\cite{PhysRevLett.119.157201,PhysRevX.8.031028}. Our study shows that YCOB is a rare KHA where the bond randomness and its influence to the QSL behavior might be precisely modelled.

\section{Conclusions And Outlook}
We have characterized the magnetic properties of large YCOB single crystals that are grown for the first time. The exchange Hamiltonian is refined by magnetic susceptibilities on the orientated single crystals. We find the bond randomness is significant, and plays an important role in the QSL behavior. No conventional magnetic transition is found, and the specific heat exhibits a nearly quadratic temperature dependence below 1 K $\sim$ 0.02$J_1$, which are phenomenologically consistent with the predictions of a gapless U(1) Dirac QSL or a $\mathbb{Z}_2$ QSL with a gap less than 0.025$J_1$.

The exact GS nature of YCOB might be further identified by the Knight shift in nuclear magnetic resonance, and by measuring inelastic neutron scattering and muon spin relaxation spectra. Future investigations
of spin dynamics and correlations of YCOB are warranted and made possible through the availability of high-quality single crystals.

\begin{acknowledgments}
We thank Tao Xiang, X. Hong, and C. Hess for helpful discussion and J. Schnack for providing their FLD data. The authors were supported by the National Key Research and Development Project of China, grant 2017YFA0302901, the Youth Innovation Promotion Association CAS, grant 2021004, and the Strategic Priority Research Program of Chinese Academy of Sciences, grant XDB33000000. This work was supported by the Fundamental Research Funds for the Central Universities, HUST: 2020kfyXJJS054.
\end{acknowledgments}

\appendix

\section{Single crystal growth, structural, thermodynamic, and electron spin resonance characterizations}
\label{sec1}

\begin{figure}
\includegraphics[width=8.6cm,angle=0]{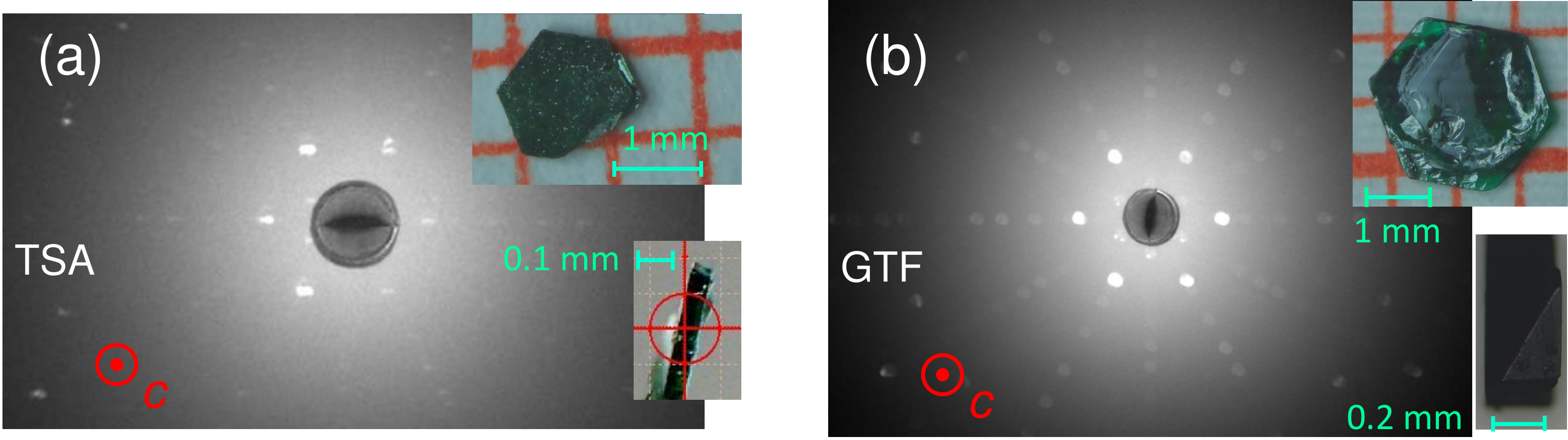}
\caption{Laue XRD patterns on the $ab$ plane of (a) the largest TSA (the upper inset) and (b) typical GTF (the upper inset) crystals. The thicknesses of typical TSA and GTF crystals are shown in the lower insets of (a) and (b), respectively.}
\label{figs1}
\end{figure}

We grew single crystals of YCu$_3$(OH)$_{6.5}$Br$_{2.5}$ (YCH) and YCu$_3$(OD$_{0.6}$H$_{0.4}$)$_{6.5}$Br$_{2.5}$ (YCD) in two ways. At first, we synthesized the crystals in a Teflon-lined stainless steel autoclave (TSA) using the hydrothermal technique~\cite{CHEN2020167066} reported by Chen et al. The high-purity D$_2$O ($\geq$ 99.9\%, Shanghai Titan Scientific Co., Ltd) was used instead of H$_2$O when we synthesized YCD. Crystals with a maximum size of 1.2$\times$1.2$\times$0.15 mm (mass $\sim$ 0.9 mg) were obtained, as shown in Fig.~\ref{figs1}(a).

To improve the quality and size of the single crystals, we performed a recrystallization process in a two-zone gradient tube furnace (GTF) with a temperature gradient of $\sim$ 2 $^{\circ}$C/cm. The starting materials of Cu(NO$_3$)$_2$$\cdot$3H$_2$O (2.416 g, 10 mmol), Y(NO$_3$)$_3$$\cdot$6H$_2$O (7.662 g, 20 mmol), KBr (7.140 g, 60 mmol), and 4 mL D$_2$O were mixed together and charged into a fused quartz tube (inside diameter 14 mm, outside diameter 20 mm, a length of $\sim$ 30 cm). The tube was sealed, mounted vertically, and then heated to 230 $^{\circ}$C in a box furnace. The temperature was maintained for 3 days and then decreased to room temperatures. The pre-reacted tube was mounted horizontally into the two-zone gradient furnace. Both the hot and cold ends were simultaneously heated to 225 and 165 $^{\circ}$C, respectively, for 6 hours. The furnace and all growth parameters were left undisturbed for 30 days, and then both ends were simultaneously cooled down to room temperatures. Transparent green single crystals with shining surfaces and a typical size of 3.1$\times$2.4$\times$0.25 mm (mass $\sim$ 7.4 mg) were obtained [see Fig.~\ref{figs1}(b)]. Back-scattering Laue x-ray diffraction (XRD) measurements (LAUESYS\_V\_674, Photonic Science \& Engineering Ltd) were carried out for both TSA and GTF single crystals. The GTF crystals show obviously sharper Laue photographs [please compare Fig.~\ref{figs1}(b) to Fig.~\ref{figs1}(a)], implying that the quality of the single crystals has been considerably improved by the above recrystallization process.

The crystal structure was obtained from the refinements of the XRD (Mo $K_{\alpha}$, $\lambda$ = 0.71073 {\AA}, XtaLAB mini \uppercase\expandafter{\romannumeral2}, Rigaku Corporation) data measured on selected YCH and YCD crystals with proper sizes. We started with the crystal structure previously reported in Ref.~\cite{CHEN2020167066}, and got a very similar refined structure (see Table~\ref{table1}). From the refinements we find that YCH(D) has a structural composition YCu$_3$[OH(D)]$_6$Br$_2$[OH(D)]$_{1-x'}$Br$_{x'}$ with $x'$ = 0.43--0.47. In comparison, a slightly larger value $x'$ = 0.51 had been reported in Ref.~\cite{CHEN2020167066}. Henceforth, we use the approximate composition YCu$_3$[OH(D)]$_{6.5}$Br$_{2.5}$ throughout this paper.

\begin{figure}
\begin{center}
\includegraphics[width=8.6cm,angle=0]{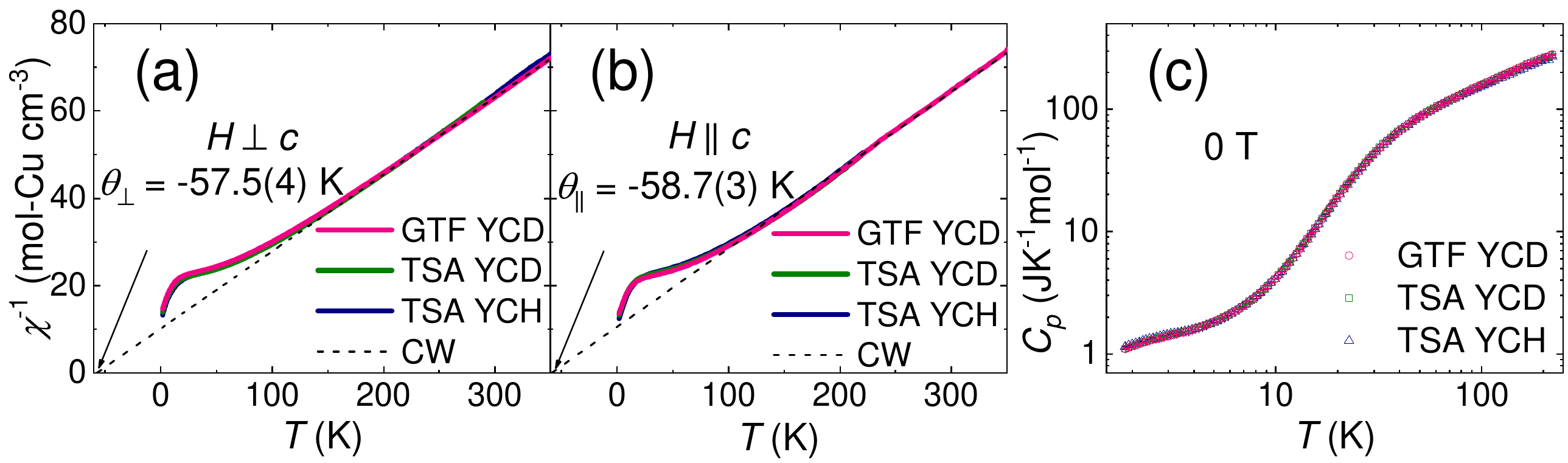}
\caption{Inverse dc susceptibilities ($H/M$) of aligned GTF single crystals of YCD (total mass 21.8 mg) measured at 1 T applied perpendicular (a) and parallel (b) to the $c$ axis, with those of TSA crystals of YCD (net weight: 23.0 mg) and YCH (net weight: 23.8 mg) for comparison. The dashed lines present the linear (Curie-Weiss) fits to the GTF data above 150 K. (c) Specific heat of a GTF single crystal of YCD (mass 5.90 mg) measured at 0 T, with those of TSA crystals of YCD (net weight: 1.90 mg) and YCH (net weight: 1.82 mg) for comparison. No substantial sample dependence of the thermodynamic properties was observed.}
\label{figs2}
\end{center}
\end{figure}

The dc magnetization and ac susceptibilities (1.8 $\leq$ $T$ $\leq$ 350 K and 0 $\leq$ $\mu_0H$ $\leq$ 7 T) were measured by a magnetic property measurement system (MPMS, Quantum Design) using well-aligned (by Laue XRD) single-crystal samples of $\sim$ 23 mg, whereas the magnetization up to 14 T was measured by a vibrating sample magnetometer in a physical property measurement system (PPMS, Quantum Design). At high temperatures the magnetization of the compound is weak, and we carefully chose a quartz holder without evident background. As shown in Fig.~\ref{figs2}(a) and \ref{figs2}(b), our susceptibilities follow the Curie-Weiss law between $\sim$ 120 and 350 K (the maximum measured temperature), with the Curie-Weiss temperatures $\theta_{\perp}$ = $-57.5(4)$ K and $\theta_{\parallel}$ = $-58.7(3)$ K along the $ab$ plane and $c$ axis, respectively. In comparison, Chen et al. reported a slightly larger magnitude of the Curie-Weiss temperature of $-74$ K from fitting their isotropic susceptibility measured in diamagnetic gelatine capsules below $\sim$ 280 K~\cite{CHEN2020167066}.

The specific heat (1.8 $\leq$ $T$ $\leq$ 220 K) was measured at a magnetic field of 0 $\leq$ $\mu_0H_{\parallel}$ $\leq$ 9 T applied along the $c$ axis using the single-crystal samples in a PPMS. The specific heat measurements on a selected GTF single crystal (mass 4.00 mg) between 0.36 and 2 K were conducted in a $^3$He refrigerator. N-grease was used to facilitate thermal contact between the crystal and the puck, and the sample coupling was better than 98\%. The contributions of the grease and puck under different applied fields were measured independently and subtracted from the data.

\begin{table*}
\caption{Structure refinements from single-crystal XRD data measured on selected YCH and YCD crystals at 300 K.}\label{table1}
\begin{center}
\begin{tabular}{ l || l | l || l | l }
    \hline
    \hline
crystals & YCH (\#1) & YCH (\#2) & YCD (\#1) & YCD (\#2) \\
    \hline
   size (mm) & 0.22 $\times$ 0.22 $\times$ 0.08 & 0.19 $\times$ 0.19 $\times$ 0.08 & 0.21 $\times$ 0.21 $\times$ 0.04 & 0.11 $\times$ 0.11 $\times$ 0.06 \\
   molar mass (g/mol) & 588.0(1.3) & 586.8(1.3) &  589.5(1.3) & 590.7(1.3) \\
   space group & $P$\={3}$m$1 (No. 164) & $P$\={3}$m$1 (No. 164) & $P$\={3}$m$1 (No. 164) & $P$\={3}$m$1 (No. 164) \\
   $a$ (= $b$, {\AA}) & 6.6647(3) & 6.6704(5) & 6.6576(4) & 6.667(1) \\
   $c$ ({\AA}) & 6.0093(3) &  6.0014(4) & 6.0151(3) & 6.004(1) \\
   cell volume ({\AA}$^3$), $Z$ = 1 & 231.16(3) &  231.26(5) & 230.89(4) & 231.1(1)  \\
    \hline
   Y1 (1$b$): occupancy $f_{\mathrm{Y}1}$ & 0.30(2) & 0.30(2) & 0.30(2) & 0.30(2) \\
   100$\times U_{iso}$ & 0.8(2) & 1.4(2) & 0.9(2) & 0.4(2) \\
   Y2 (2$c$): occupancy $f_{\mathrm{Y}2}$ & 0.35(1) & 0.35(1) & 0.35(1) & 0.35(1) \\
   $z$ & 0.374(2) & 0.374(2) & 0.373(2) & 0.373(2) \\
   100$\times U_{iso}$ & 0.8(2) & 1.4(2) & 0.9(2) & 0.4(2) \\
    \hline
   Cu (3$f$, occupancy = 1): 100*$U_{iso}$ & 0.8(1) & 1.7(1) & 0.8(1) & 0.3(1) \\
    \hline
   Br1 (2$d$, occupancy = 1): $z$ & 0.8566(6) & 0.8566(6) & 0.8560(5) & 0.8566(7) \\
   100$\times U_{iso}$ & 1.8(1) & 2.6(1) & 1.8(1) & 1.4(1) \\
    \hline
   O1 (6$i$, occupancy = 1): $x$ ($x$ = 0.5$y$) & 0.1896(9) & 0.1885(9) & 0.1894(8) & 0.189(1) \\
   $z$ & 0.372(3) & 0.362(2) & 0.366(2) & 0.378(3) \\
   100$\times U_{iso}$ & 1.0(3) & 1.4(4) & 1.0(3) & 1.5(5) \\
    \hline
    Br2 (1$a$): occupancy $f_{\mathrm{Br}2}$ & 0.47(2) & 0.45(2) & 0.43(2) & 0.45(2) \\
   100$\times U_{iso}$ & 2.3(2) & 2.8(4) & 2.5(3) & 1.6(4) \\
   O2 (1$a$): occupancy $f_{\mathrm{O}2}$ & 0.53(2) & 0.55(2) & 0.57(2) & 0.55(2) \\
   100$\times U_{iso}$ & 2.3(2) & 2.8(4) & 2.5(3) & 1.6(4) \\
    \hline
   $h$ range & $-9 \rightarrow 9$ & $-9 \rightarrow 9$ & $-8 \rightarrow 9$ & $-7 \rightarrow 8$ \\
   $k$ range & $-8 \rightarrow 11$ & $-8 \rightarrow 11$ & $-9 \rightarrow 11$ & $-9 \rightarrow 11$ \\
   $l$ range & $-8 \rightarrow 10$ & $-3 \rightarrow 10$ & $-8 \rightarrow 10$ & $-8 \rightarrow 10$ \\
   reflections ($I$ $>$ 0) & 954 & 795  & 841  & 560 \\
   reflections ($I$ $>$ 3$\sigma$($I$)) & 837 & 674  & 740  & 492 \\
   $R$($F$) ($I$ $>$ 3$\sigma$($I$)) & 3.8\% & 6.7\% & 7.1\% & 9.7\% \\
   $R_{w}$($F$) ($I$ $>$ 3$\sigma$($I$)) & 3.3\% & 4.8\% & 6.1\% & 6.6\% \\
    \hline
    \hline
\end{tabular}
\end{center}
\end{table*}

No significant sample dependence of thermodynamic properties was observed among GTF YCD, TSA YCD, and TSA YCH single crystals (see Fig.~\ref{figs2}). Therefore, we focused on the measurements on the highly-quality GTF YCD (mentioned as YCOB) single crystals below and also throughout the main text.

The YCOB crystals are good insulators and have resistance larger than 20 M$\Omega$ measured parallel and perpendicular to the $c$ axis at room temperatures. Above $\sim$ 120 K, the inverse susceptibilities of YCOB show a linear temperature dependence following the Curie-Weiss law along both the $ab$ plane and $c$ axis [see Fig.~\ref{figs2}(a) and \ref{figs2}(b)]. We obtain the lengths of the magnetic moments $g_{\perp}\mu_{\mathrm{B}}\sqrt{S(S+1)}$ = $\sqrt{3k_\mathrm{B}C_{\perp}/(N_\mathrm{A}\mu_0)}$ = 1.898(1)$\mu_{\mathrm{B}}$ and $g_{\parallel}\mu_{\mathrm{B}}\sqrt{S(S+1)}$ = $\sqrt{3k_\mathrm{B}C_{\parallel}/(N_\mathrm{A}\mu_0)}$ = 1.882(1)$\mu_{\mathrm{B}}$ from the fitted Curie constants $C_{\perp}$ = 5.661(7) Kcm$^3$ per mol Cu and $C_{\parallel}$ = 5.568(4) Kcm$^3$ per mol Cu, respectively. By fixing $S$ = $\frac{1}{2}$, we further get $g_{\perp}$ = 2.192(1) and $g_{\parallel}$ = 2.174(1). Moreover, the antiferromagnetic exchange energy can be roughly estimated as $J_1$ $\sim$ $-3\theta_{\perp}/$[4$S$($S$+1)] $\sim$ $-3\theta_{\parallel}/$[4$S$($S$+1)] $\sim$ 60 K.

\begin{figure}
\begin{center}
\includegraphics[width=8.6cm,angle=0]{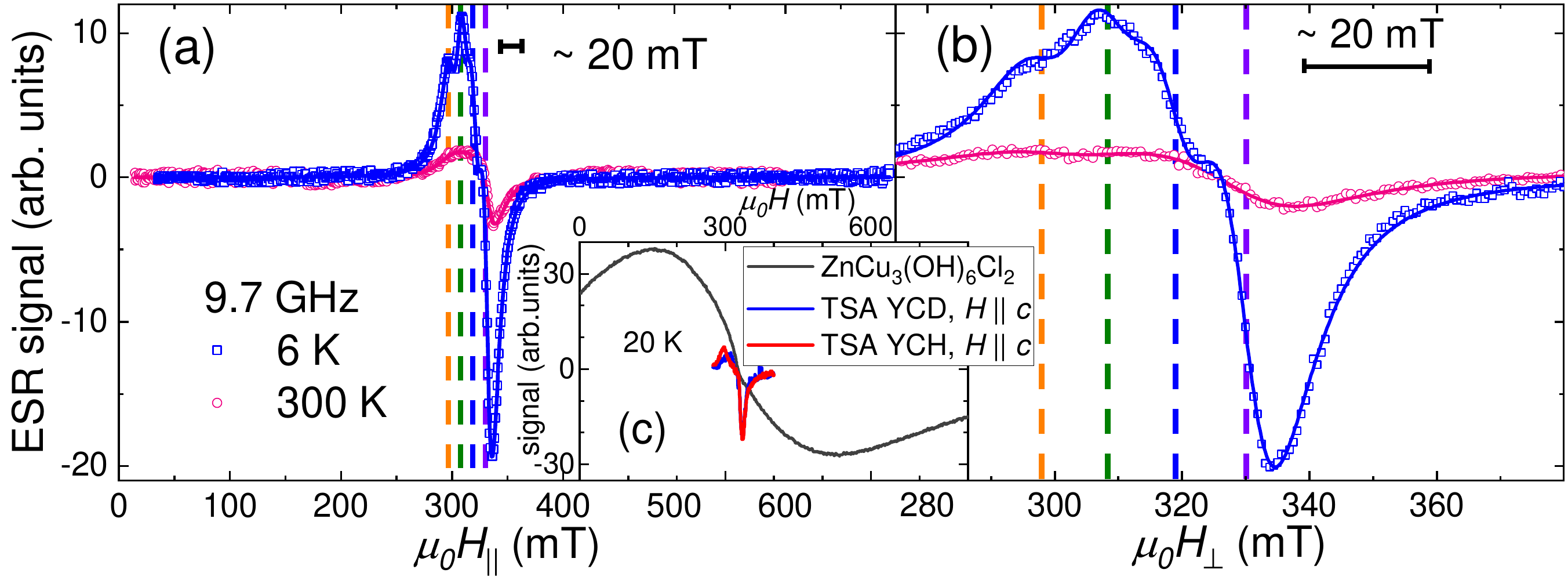}
\caption{The ESR spectra measured parallel (a) and perpendicular (b) to the $c$ axis. The solid lines show the four-Lorentzian fits [see Eq.~(\ref{eqS2})], and the dashed lines present the calculated ESR fields (see Fig.~\ref{figs4}). (c) The x-band ESR spectra measured at 20 K on single-crystal samples of TSA YCH (23.8 mg) and YCD (23.0 mg), with the spectrum of the ZnCu$_3$(OH)$_6$Cl$_2$ powder ($\sim$ 20 mg) for comparison.}
\label{figs3}
\end{center}
\end{figure}

\begin{figure*}
\begin{center}
\includegraphics[width=17cm,angle=0]{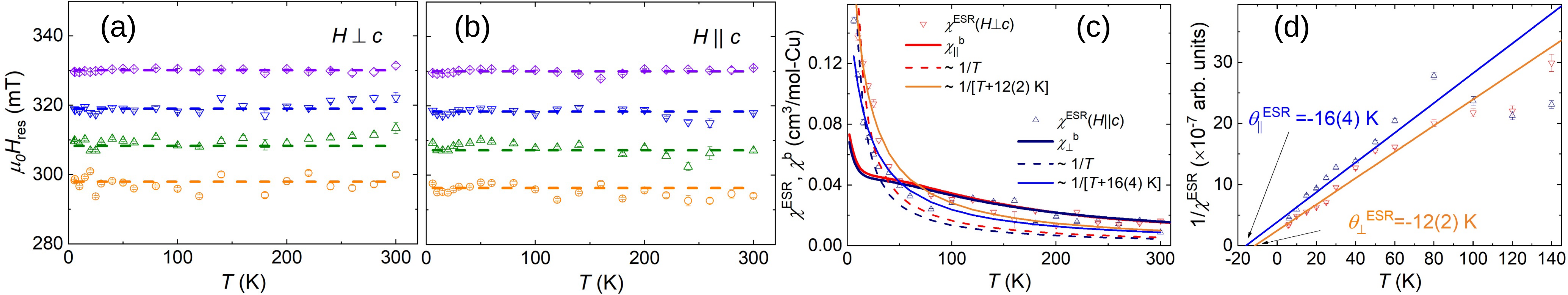}
\caption{ESR fields obtained from fitting the ESR spectra measured along the $ab$ plane (a) and $c$ axis (b), respectively (see Fig.~\ref{figs3}). The dashed lines show the combined fit. (c) Scaling of the ESR intensities, $\chi^{\mathrm{ESR}}(H\perp c)$ and $\chi^{\mathrm{ESR}}(H\parallel c)$, with the bulk susceptibilities (thick lines), $\chi^{\mathrm{b}}_{\parallel}$ and $\chi^{\mathrm{b}}_{\perp}$, respectively. The thin dashed and solid lines present the Curie-Weiss fits with zero (i.e., Curie tails) and adjustable Curie-Weiss temperatures, respectively. (d) Inverse (integrated) ESR intensities.}
\label{figs4}
\end{center}
\end{figure*}

The first derivative electron spin resonance (ESR) absorption spectra were measured at 6--300 K on the aligned single-crystal samples of YCOB (total mass $\sim$ 22 mg) with the magnetic field applied parallel and perpendicular to the $c$ axis, respectively, using continuous wave spectrometers (CIQTEK EPR200-Plus and Bruker EMXmicro-6/1) at x-band frequencies ($\nu$ $\sim$ 9.7 GHz) equipped with $^4$He refrigerators. The single-crystal samples of YCOB were washed with high-purity water and ethanol repeatedly and in succession, and the absence of any obvious impurity phase was confirmed by using a microscope, before the ESR measurements. The ESR spectra measured at various temperatures are so narrow that the distinctive hyperfine structure is clearly visible (see Fig.~\ref{figs3})~\cite{abragam2012electron,1990ESR}. The similar ESR lines usually appear in Cu$^{2+}$-based complexes with weak couplings~\cite{1990ESR}. However, in most of the existing strongly-correlated 3$d$ magnets, e.g., ZnCu$_3$(OH)$_6$Cl$_2$~\cite{PhysRevLett.101.026405} and $\alpha$-CrOOH~\cite{Liu2021Frustrated}, broad single ESR lines are typically observed due to the magnetic anisotropy. As shown in Fig.~\ref{figs3}, the ESR signals are still robust even at room temperatures, with the maximum strength of $\sim$ 1 ($\mu$V) at the microwave power of 0.2 mW, modulation amplitude of 0.4 mT, and magnification of 10. The integrated intensity of YCOB is about 1\% of that of ZnCu$_3$(OH)$_6$Cl$_2$ reference powder, roughly consistent with the estimated concentration ($\sim$ 0.8\%) of the local moments [see Fig.~\ref{figs3} (c)]. Moreover, the integrated ESR intensity roughly follows a Curie-Weiss behavior with the Curie-Weiss temperature of $\sim$ $-$14 K, instead of the bulk susceptibility [see Figs.~\ref{figs4} (c) and \ref{figs4}(d)]. These observations suggest that the narrow ESR signals of YCOB should mainly originate from the nearly free local moments induced by the bond randomness on the kagome lattice (see main text). In contrast, the sibling YCu$_3$(OH)$_6$Cl$_3$ without evident bond randomness shows a very broad ESR signal with the linewidth of $\sim$ 7 T~\cite{PhysRevLett.125.027203}. Therefore, the possible broad ESR lines from the strongly-coupled ($J_1$ $\sim$ 60 K) Cu spins of YCOB may not be well captured by our x-band (maximum applied field of $\sim$ 1 T) ESR spectrometers.

The single-ion hyperfine Hamiltonian of the Cu$^{2+}$ ($S$ = $\frac{1}{2}$, $I$ = $\frac{3}{2}$) ions that is invariant under the $D_{3d}$ point group symmetry can be written as~\cite{abragam2012electron,1990ESR}
\begin{widetext}
\begin{multline}
\mathcal{H}_{\mathrm{hp}}=-\mu_0\mu_\mathrm{B}\mathbf{H}\cdot\mathbf{g}\cdot\mathbf{S}+\mathbf{I}\cdot\mathbf{A}\cdot\mathbf{S}\\
\tiny
=\left(
\begin{array}{cccccccc}
\frac{3}{4}A_{\parallel}-\frac{1}{2}G_{\parallel}&0&0&0&\frac{i}{2}G_{\perp}&0&0&0\\
0&\frac{1}{4}A_{\parallel}-\frac{1}{2}G_{\parallel}&0&0&\frac{\sqrt{3}}{2}A_{\perp}&\frac{i}{2}G_{\perp}&0&0\\
0&0&-\frac{1}{4}A_{\parallel}-\frac{1}{2}G_{\parallel}&0&0&A_{\perp}&\frac{i}{2}G_{\perp}&0\\
0&0&0&-\frac{3}{4}A_{\parallel}-\frac{1}{2}G_{\parallel}&0&0&\frac{\sqrt{3}}{2}A_{\perp}&\frac{i}{2}G_{\perp}\\
-\frac{i}{2}G_{\perp}&\frac{\sqrt{3}}{2}A_{\perp}&0&0&-\frac{3}{4}A_{\parallel}+\frac{1}{2}G_{\parallel}&0&0&0\\
0&-\frac{i}{2}G_{\perp}&A_{\perp}&0&0&-\frac{1}{4}A_{\parallel}+\frac{1}{2}G_{\parallel}&0&0\\
0&0&-\frac{i}{2}G_{\perp}&\frac{\sqrt{3}}{2}A_{\perp}&0&0&\frac{1}{4}A_{\parallel}+\frac{1}{2}G_{\parallel}&0\\
0&0&0&-\frac{i}{2}G_{\perp}&0&0&0&\frac{3}{4}A_{\parallel}+\frac{1}{2}G_{\parallel}\\
\end{array}\right),
\label{eqS1}
\end{multline}
\end{widetext}
\normalsize
under the subspace of $|S^z=\frac{1}{2},I^z=\frac{3}{2}\rangle,~|\frac{1}{2},\frac{1}{2}\rangle,~|\frac{1}{2},-\frac{1}{2}\rangle,~|\frac{1}{2},-\frac{3}{2}\rangle,~|-\frac{1}{2},\frac{3}{2}\rangle,~|-\frac{1}{2},\frac{1}{2}\rangle,~|-\frac{1}{2},-\frac{1}{2}\rangle,~|-\frac{1}{2},-\frac{3}{2}\rangle$. Here, $\mathbf{H}$ = $H$[0 $\sin\theta$ $\cos\theta$] is the applied magnetic field, $G_{\parallel}$ = $\mu_0\mu_\mathrm{B}Hg_{\parallel}\cos\theta$ and $G_{\perp}$ = $\mu_0\mu_\mathrm{B}Hg_{\perp}\sin\theta$ are the Zeeman terms. The eight eigenstates of Supplementary Eq.~(\ref{eqS1}) are represented as $|E^{\eta}_j\rangle$ ($j$ = 1, 2, ..., 8) with increasing eigenenergies. Due to $A_{\eta}$ $\ll$ $\mu_0\mu_\mathrm{B}H_{\mathrm{res}}^{\eta}g_{\eta}$, these eigenstates give rise to four ESR modes with $E^{\eta}_{9-j'}-E^{\eta}_{j'}$ = $h\nu$ and nonzero transition probabilities $\propto$ $|\langle E^{\eta}_{9-j'}|2S^x|E^{\eta}_{j'}\rangle|^2$ $\sim$ 1, where $j'$ = 1, 2, 3, 4. These four ESR modes are governed by the transitions, $|\frac{1}{2},I^z\rangle\rightarrow|-\frac{1}{2},I^z\rangle$ ($I^z$ = $-\frac{3}{2}$, $-\frac{1}{2}$, $\frac{1}{2}$, $\frac{3}{2}$), similar to other Cu$^{2+}$-based magnets~\cite{1990ESR}. Therefore, we fit the experimental ESR spectra (see Fig.~\ref{figs3}) using,
\begin{equation}
\frac{dI_{\mathrm{abs}}^{\eta}}{\mu_0dH}=\frac{16I_0^{\eta}\Delta H_{\eta}}{\pi\mu_0^2}\sum_{j' = 1}^{4}\frac{f_{j'}^{\eta}(H_{j'}^{\eta}-H)}{[4(H_{j'}^{\eta}-H)^2+\Delta H_{\eta}^2]^2},
\label{eqS2}
\end{equation}
where $f_{j'}^{\eta}$ $>$ 0 is the spectrum weight of each ESR mode with $\sum_{j' = 1}^{4}f_{j'}^{\eta}$ $\equiv$ 1, $\mu_0\Delta H_{\eta}$ is the line width that is presumed to be independent of $j'$ for simplicity, and $I_0^{\eta}$ is the total integrated intensity that is proportional to the dynamical susceptibility in the direction perpendicular to the applied field, $I_0^{\eta}$ $\propto$ $\chi_{\eta'}''$(\textbf{q} $\rightarrow$ 0, $\omega$)~\cite{PhysRevLett.101.026405}. The fitted ESR fields are almost temperature-independent [see Figs.~\ref{figs4}(a) and \ref{figs4}(b)]. From fitting the ESR fields both parallel ($\eta$ = $\parallel$) and perpendicular ($\eta$ = $\perp$) to the $c$ axis, we obtain all the single-ion parameters, $g_{\perp}$ = 2.205(4), $g_{\parallel}$ = 2.211(4), $|A_{\perp}|$ = 15.8(2) mK and $|A_{\parallel}|$ = 16.7(2) mK.

\section{DFT + $U$ calculations}
\label{sec2}

\begin{figure*}
\begin{center}
\includegraphics[width=16.5cm,angle=0]{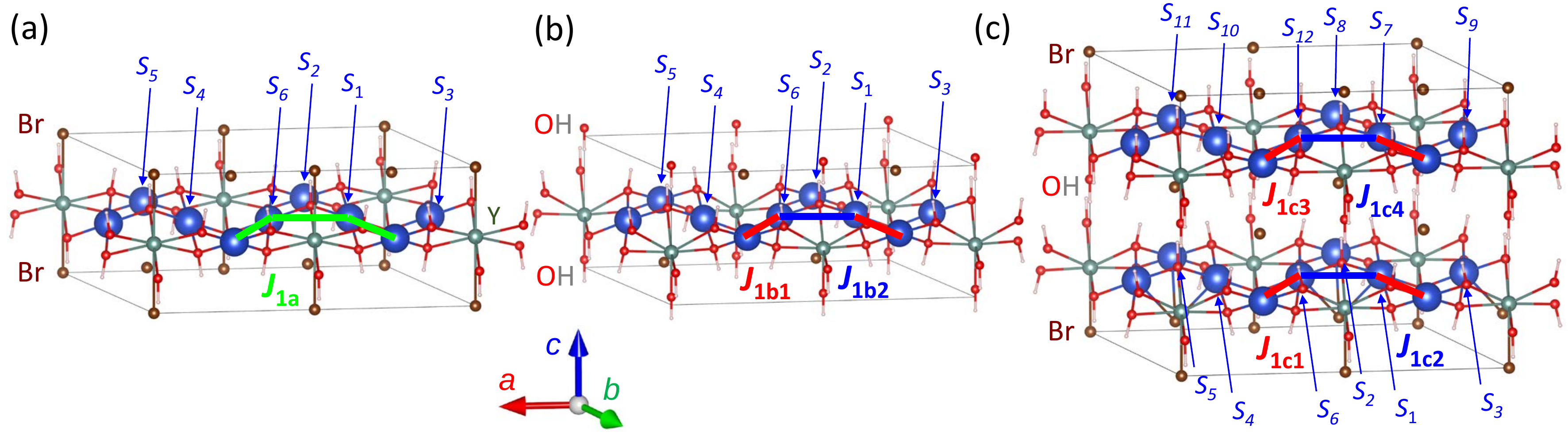}
\caption{Optimized crystal structures for different stacking sequences of Br/OH, (a) symmetric Br-Br [2YCu$_3$(OH)$_6$Br$_3$, supercell: 2$\times$1$\times$1 unit cells, total energy $-$31.2 eV/Cu], nonsymmetric (b) OH-OH [2YCu$_3$(OH)$_7$Br$_2$, 2$\times$1$\times$1, total energy $-$33.9 eV/Cu] and (c) Br-OH-Br [4YCu$_3$(OH)$_{6.5}$Br$_{2.5}$, 2$\times$1$\times$2, total energy $-$32.5 eV/Cu]. The indices of Cu$^{2+}$ (blue ball) spins and superexchanges are marked. It is worth to note that the symmetric OH-HO configuration (total energy $-$33.6 eV/Cu) isn't energetically favorable compared to the nonsymmetric one (see b, total energy $-$33.9 eV/Cu).}
\label{figv1}
\end{center}
\end{figure*}

Although our single-crystal XRD measurements (see above), as well as the reported electron-microprobe analysis~\cite{CHEN2020167066}, show obvious site-mixing disorder of OH$^{-}$ and Br$^{-}$, neither global symmetry reduction nor antisite mixing between Cu$^{2+}$ and other nonmagnetic ions is probed in YCOB. To go beyond the average structure from single-crystal XRD and investigate the influence of this structural disorder on the magnetism, we conducted the density functional theory (DFT) + $U$ simulation, which had been widely used in other Cu$^{2+}$-based magnets, e.g., volborthite Cu$_3$V$_2$O$_7$(OH)$_2$$\cdot$2H$_2$O~\cite{PhysRevLett.117.037206}, kapellasite Cu$_3$Zn(OH)$_6$Cl$_2$, and haydeeite Cu$_3$Mg(OH)$_6$Cl$_2$~\cite{PhysRevLett.101.106403}. Similar to YbMgGaO$_4$~\cite{PhysRevLett.118.107202}, we constructed three OH/Br configurations to simulate the site mixing, and optimized their geometry using DFT calculations (see Fig.~\ref{figv1}). All possible nearest-neighbor (NN) environments of 1$b$ position are included in the three configurations (see Fig.~\ref{figv1}), and the further enlarging of supercells sharply increases the computational cost. All our DFT + $U$ calculations were performed in the VASP code~\cite{vasp1,vasp2} with the generalized gradient approximation (GGA)~\cite{PhysRevLett.77.3865}. The 6$\times$6$\times$6 k-mesh was used, and residual forces were below 0.006 eV/{\AA} in the fully optimized structures. The lattice parameters were fixed to the experimental values, $a$ = $b$ = 6.6647 {\AA} and $c$ = 6.0093 {\AA}.

The random distribution of OH$^{-}$ and Br$^{-}$ takes place around the 1$a$ positions, which are adjacent to the hexagons on the kagome lattice of magnetic Cu$^{2+}$ ions (see Fig.~\ref{figv1}). The polar OH$^{-}$ ions easily make the local environments of 1$b$ positions (i.e., the Y1 sites, located at the centers of the hexagons) nonsymmetric, and drive Y$^{3+}$ away from its ideal 1$b$ position, i.e., $|\Delta z_{\mathrm{Y}2}|$ $>$ 0. Our DFT calculation gives $|\Delta z_{\mathrm{Y}1}|$ = 0 in the symmetric Br-Br [see Fig.~\ref{figv1} (a)] stacking, and $|\Delta z_{\mathrm{Y}2}|$ $\sim$ 0.6 and 0.4 {\AA} in the nonsymmetric OH-OH [see Fig.~\ref{figv1} (b)] and Br-OH-Br [see Fig.~\ref{figv1} (c)] cases, respectively, which are well consistent with the XRD results of $|\Delta z_{\mathrm{Y}1}|$ = 0  and $|\Delta z_{\mathrm{Y}2}|$ $\sim$ 0.7 {\AA} (Table ~\ref{table1}). Furthermore, we find that both $\angle$CuOCu and Cu-O, which are closely relevant to the exchange integrals, are sensitively dependent on the OH/Br distribution (see Table ~\ref{table2}). When the local environment of the 1$b$ Wyckoff position gets nonsymmetric, the configuration of two alternate superexchanges along the hexagon is naturally expected, according to the structural symmetry [see Figs.~\ref{figv1} (b) and \ref{figv1}(c)].

\begin{table*}
\caption{Optimized bond angles of $\angle$CuOCu and bond lengths of Cu-O for different stacking sequences, as well as the NN exchange couplings calculated by DFT + $U$ (see Table~\ref{table3}).}\label{table2}
\begin{center}
\begin{tabular}{ l || l | l | l || l | l | l || l | l | l || l | l | l }
    \hline
    \hline
supercells & $\angle$CuOCu & Cu-O & coupling & $\angle$CuOCu & Cu-O & coupling & $\angle$CuOCu & Cu-O & coupling & $\angle$CuOCu & Cu-O & coupling \\
    \hline
  Br-Br & 115.5$^\circ$ & 1.97{\AA} & $J_{1\mathrm{a}}$=56K & $-$ & $-$ & $-$ & $-$ & $-$ & $-$ & $-$ & $-$ & $-$ \\
  OH-OH & 116.0$^\circ$ & 1.97{\AA} & $J_{1\mathrm{b}1}$=94K & 112.7$^\circ$ & 1.99{\AA} & $J_{1\mathrm{b}2}$=38K & $-$ & $-$ & $-$ & $-$ & $-$ & $-$ \\
  Br-OH-Br & 116.6$^\circ$ & 1.99{\AA} & $J_{1\mathrm{c}1}$=93K & 108.2$^\circ$ & 2.02{\AA} & $J_{1\mathrm{c}2}=-$11K & 117.0$^\circ$ & 1.95{\AA} & $J_{1\mathrm{c}3}$=101K & 116.8$^\circ$ & 1.96{\AA} & $J_{1\mathrm{c}4}$=95K \\
    \hline
    \hline
\end{tabular}
\end{center}
\end{table*}

\begin{table*}
\caption{Collinear magnetic states and their energy differences from the fully ferromagnetic state. From fitting these energy differences, we obtain the exchange couplings, including NN $J_1$, second-neighbor $J_2$, and interlayer $J_{\perp}$. The least-squares standard deviation is defined as $\sigma$ = $\sqrt{\frac{1}{N_0}\sum_i(X_i^{\mathrm{DFT}}-X_i^{\mathrm{fit}})^2}$, where $X_i^{\mathrm{DFT}}$ and $X_i^{\mathrm{fit}}$ are the DFT and fitted values, respectively, $N_0$ is the number of the data points.}\label{table3}
\begin{center}
\begin{tabular}{ l || l | l | l | l | l | l | l }
    \hline
    \hline
Br-Br, $St$ = $\mid S_1^zS_2^zS_3^zS_4^zS_5^zS_6^z\rangle$ & $\mid\downarrow\uparrow\uparrow\downarrow\uparrow\uparrow\rangle$ & $\mid\downarrow\downarrow\uparrow\downarrow\uparrow\uparrow\rangle$ & $\mid\downarrow\uparrow\uparrow\uparrow\uparrow\uparrow\rangle$ & $\mid\uparrow\downarrow\uparrow\downarrow\uparrow\uparrow\rangle$ & $\mid\uparrow\downarrow\downarrow\downarrow\uparrow\uparrow\rangle$ & $\mid\uparrow\downarrow\downarrow\downarrow\downarrow\uparrow\rangle$ & $\mid\uparrow\uparrow\downarrow\uparrow\uparrow\uparrow\rangle$ \\
    \hline
DFT + $U$, $E$($\mid\uparrow\uparrow\uparrow\uparrow\uparrow\uparrow\rangle$)$-E$($St$) (K) & 232.7 & 234.4 & 117.7 & 231.4 & 229.3 & 174.4 & 116.9 \\
least-squares fitted $E$($\mid\uparrow\uparrow\uparrow\uparrow\uparrow\uparrow\rangle$)$-E$($St$) (K) & 233.2 & 234.1 & 117.5 & 230.8 & 229.9 & 174.9 & 115.7 \\
    \hline
exchange couplings and standard deviation & \multicolumn{7}{|c}{$J_{1\mathrm{a}}$ = 56.2 K, $J_{2\mathrm{a}}$ = 2.0 K, $J_{3\mathrm{a}}$ = $-$0.4 K, $J_{d\mathrm{a}}$ = 1.2 K, and $\sigma_\mathrm{a}$ = 0.6 K} \\
    \hline
\end{tabular}
\begin{tabular}{ l || l | l | l | l | l | l | l }
    \hline
OH-OH, $St$ = $\mid S_1^zS_2^zS_3^zS_4^zS_5^zS_6^z\rangle$ & $\mid\downarrow\uparrow\uparrow\downarrow\uparrow\uparrow\rangle$ & $\mid\downarrow\downarrow\uparrow\downarrow\uparrow\uparrow\rangle$ & $\mid\downarrow\uparrow\uparrow\uparrow\uparrow\uparrow\rangle$ & $\mid\uparrow\downarrow\uparrow\downarrow\uparrow\uparrow\rangle$ & $\mid\uparrow\downarrow\downarrow\downarrow\uparrow\uparrow\rangle$ & $\mid\uparrow\downarrow\downarrow\downarrow\downarrow\uparrow\rangle$ & $\mid\uparrow\uparrow\downarrow\uparrow\uparrow\uparrow\rangle$ \\
    \hline
DFT + $U$, $E$($\mid\uparrow\uparrow\uparrow\uparrow\uparrow\uparrow\rangle$)$-E$($St$) (K) & 265.7 & 264.4 & 132.0 & 263.0 & 260.9 & 224.8 & 132.2 \\
least-squares fitted $E$($\mid\uparrow\uparrow\uparrow\uparrow\uparrow\uparrow\rangle$)$-E$($St$) (K) & 265.9 & 264.7 & 131.7 & 262.5 & 261.3 & 224.8 & 131.7 \\
    \hline
exchange couplings and standard deviation & \multicolumn{7}{|c}{$J_{1\mathrm{b}1}$ = 94 K, $J_{1\mathrm{b}2}$ = 38 K, $J_{2\mathrm{b}}$ = 0.5 K, $J_{3\mathrm{b}}$ = $-$0.6 K, $J_{d\mathrm{b}}$ = $-$0.6 K, and $\sigma_\mathrm{b}$ = 0.3 K} \\
    \hline
\end{tabular}
\begin{tabular}{ l || l | l | l | l | l | l | l }
    \hline
Br-OH-Br, $St$ = & $\mid\downarrow\uparrow\uparrow\downarrow\uparrow\uparrow$ & $\mid\uparrow\uparrow\uparrow\uparrow\uparrow\uparrow$ & $\mid\downarrow\uparrow\uparrow\downarrow\uparrow\uparrow$ & $\mid\downarrow\uparrow\uparrow\uparrow\downarrow\uparrow$ & $\mid\uparrow\uparrow\downarrow\downarrow\downarrow\uparrow$ & $\mid\uparrow\uparrow\downarrow\downarrow\downarrow\uparrow$ & $\mid\uparrow\uparrow\uparrow\uparrow\uparrow\uparrow$ \\
$\mid S_1^zS_2^zS_3^zS_4^zS_5^zS_6^zS_7^zS_8^zS_9^zS_{10}^zS_{11}^zS_{12}^z\rangle$ & $\uparrow\uparrow\uparrow\uparrow\uparrow\uparrow\rangle$ & $\downarrow\uparrow\uparrow\downarrow\uparrow\uparrow\rangle$ & $\uparrow\downarrow\downarrow\uparrow\downarrow\downarrow\rangle$ & $\uparrow\downarrow\uparrow\downarrow\uparrow\uparrow\rangle$ & $\uparrow\uparrow\downarrow\downarrow\downarrow\uparrow\rangle$ & $\uparrow\uparrow\uparrow\uparrow\uparrow\uparrow\rangle$ & $\uparrow\uparrow\downarrow\downarrow\downarrow\uparrow\rangle$ \\
    \hline
DFT + $U$, $E$($\mid\uparrow\uparrow\uparrow\uparrow\uparrow\uparrow\uparrow\uparrow\uparrow\uparrow\uparrow\uparrow\rangle$)$-E$($St$) (K) & 172.7 & 400.4 & 570.3 & 565.1 & 398.6 & 194.6 & 209.5 \\
least-squares fitted $E$($\mid\uparrow\uparrow\uparrow\uparrow\uparrow\uparrow\uparrow\uparrow\uparrow\uparrow\uparrow\uparrow\rangle$)$-E$($St$) (K) & 171.3 & 399.0 & 571.7 & 565.1 & 399.0 & 194.1 & 209.0 \\
    \hline
exchange couplings and standard deviation & \multicolumn{7}{|c}{$J_{1\mathrm{c}1}$ = 93 K, $J_{1\mathrm{c}2}$ = $-$11 K, $J_{1\mathrm{c}3}$ = 101 K, $J_{1\mathrm{c}4}$ = 95 K} \\
    & \multicolumn{7}{|c}{$J_{2\mathrm{c}}$ = 1.3 K, $J_{\perp\mathrm{c}}$ = 0.7 K, and $\sigma_\mathrm{c}$ = 0.9 K}\\
    \hline
    \hline
\end{tabular}
\end{center}
\end{table*}

Starting with the optimized crystal structures, we further calculated the exchange couplings from the energy differences between the fully ferromagnetic state and various collinear states. The exchange integrals monotonically decrease with the increasing of the Coulomb repulsion $U_{\mathrm{eff}}$~\cite{PhysRevLett.117.037206,PhysRevLett.101.106403,PhysRevLett.125.027203}. For YCOB, we fixed $U_{\mathrm{eff}}$ = 8.72 eV in all GGA + $U$ calculations, and obtained the average $\bar{J}_1$ = (2$J_{1\mathrm{a}}$+$J_{1\mathrm{b}1}$+$J_{1\mathrm{b}2}$+$J_{1\mathrm{c}1}$+$J_{1\mathrm{c}2}$+$J_{1\mathrm{c}3}$+$J_{1\mathrm{c}4}$)/8 = 65 K (see Table ~\ref{table3}), in good agreement with the experimental result of $|\theta_{\perp}|$ $\sim$ $|\theta_{\parallel}|$ $\sim$ 60 K. From fitting these energy differences, we obtain the exchange couplings for each stacking sequence of Br/OH (see Table ~\ref{table3}).

Based on the above DFT + $U$ calculations, the following conclusions can be drawn:

(1) In YCOB, all the long-distance couplings beyond NNs are negligible, $|J_2|$/$\bar{J}_1$ $<$ 4\%, $|J_3|$/$\bar{J}_1$ $<$ 2\%, $|J_d|$/$\bar{J}_1$ $<$ 2\%, and $|J_{\perp}|$/$\bar{J}_1$ $<$ 2\% (see Table ~\ref{table3}), due to the spatial localization of the 3$d$ electrons (highly similar to YCu$_3$(OH)$_6$Cl$_3$~\cite{PhysRevLett.125.027203}). This significantly simplifies the exchange model for YCOB.

(2) The antisite disorder of OH/Br strongly influences the NN exchange couplings ($J_1$), and causes the alternation of two NN exchanges along the randomly distributed (1$-f_{\mathrm{Y}1}$$\sim$ 70\%) nonsymmetric hexagons of the kagome lattice, ($J_{1\mathrm{c}3}-J_{1\mathrm{c}4}$)/($J_{1\mathrm{c}3}+J_{1\mathrm{c}4}$) $\sim$ 0.03, ($J_{1\mathrm{b}1}-J_{1\mathrm{b}2}$)/($J_{1\mathrm{b}1}+J_{1\mathrm{b}2}$) $\sim$ 0.43, and ($J_{1\mathrm{c}1}-J_{1\mathrm{c}2}$)/($J_{1\mathrm{c}1}+J_{1\mathrm{c}2}$) $\sim$ 1.26.

(3) There exist no orphan spins, even in the presence of the mixing between OH and Br. The antisite disorder occurs around the high-symmetry Wyckoff position, 1$a$, and thus the weaker and stronger bonds alternate along the hexagon in the nonsymmetric environment. Although one of the alternate $\angle$CuOCu may be profoundly reduced to $\sim$ 108.2$^{\circ}$ (see Table ~\ref{table2}), and the strength of the coupling can be surprisingly weak, e.g., $|J_{1\mathrm{c}2}|$ $\sim$ 11 K, there exist no orphan Cu$^{2+}$ spins that are weakly coupled to all the other ones.

Similar to the case of the nonmagnetic impurities in the kagome Heisenberg antiferromagnet (KHA)~\cite{PhysRevB.77.184423,PhysRevB.79.214415,PhysRevLett.104.177203}, nearly free local moments can nevertheless be induced by the bond randomness in YCOB, which accounts for almost all of the low-temperature observations (see main text for model no.2), including the weak upturn seen in susceptibilities, the slight polarization in magnetization, and the very narrow ESR signals (see above).

\section{Refinement of the exchange Hamiltonian.}
\label{sec3}

In this section, we make great efforts to determine the exchange Hamiltonian of the KHA YCOB. At present, we have collected both susceptibility and specific heat data of YCOB, which can be used to refine the spin Hamiltonian. However, it is still extremely challenging to extract precisely the magnetic contribution from the total specific heat above $\sim$ 10 K (see Appendix~\ref{sec4}), in the absence of nonmagnetic reference compounds. Therefore, below we mainly focus on the magnetic susceptibilities of the YCOB single crystal measured both parallel and perpendicular to the $c$ axis.

\subsection{No evident orphan spins.}
\label{sec3p0}

In YCOB, the significant antisite disorder between magnetic Cu$^{2+}$ and other nonmagnetic ions, i.e., Y$^{3+}$, H(D)$^{+}$, O$^{2-}$, Br$^-$ is expected to be prohibited owing to the large differences in ionic charges and radii, and thus orphan spins should be negligible without any obvious impurity phases~\cite{CHEN2020167066,arXiv:2107.11942}. Following the method previously used in the KHA herbertsmithite~\cite{PhysRevB.76.132411}, we try to reproduce the magnetization measured at 2 K and up to 14 T ($\mu_\mathrm{B}g_{\eta}\mu_0H$ $\sim$ 0.35$J_1$) parallel and perpendicular to the $c$ axis, by a combination of a Brillouin function for ``free" spins and a linear magnetization for ``intrinsic" kagome spins [see Fig.~\ref{fig_free}(a)],
\begin{equation}
M_{\eta}=\frac{f_{\eta}g_{\eta}N_\mathrm{A}\mu_\mathrm{B}}{2}\frac{\exp(\frac{\mu_\mathrm{B}g_{\eta}\mu_0H}{k_\mathrm{B}T})-1}{\exp(\frac{\mu_\mathrm{B}g_{\eta}\mu_0H}{k_\mathrm{B}T})+1}+\chi_i^{\eta}H,
\label{eq_free1}
\end{equation}
where $f_{\eta}$ and $\chi_i^{\eta}$ are fitting parameters for the fraction of free spins and nonsaturated ``intrinsic" susceptibilities, respectively. We obtain $f_{\parallel}$ = 0.80\%, $\chi_i^{\parallel}$ = 0.048 cm$^3$/mol-Cu and $f_{\perp}$ = 0.75\%, $\chi_i^{\perp}$ = 0.043 cm$^3$/mol-Cu parallel and perpendicular to the $c$ axis, respectively.

However, when the formalism of the free spins is extended to the specific heat, we find obvious inconsistency. The specific heat contributed by free spins is given by,
\begin{equation}
C_f(\mu_0H_{\parallel})=f_{\parallel}R\frac{[\frac{\Delta E(\mu_0H_{\parallel})}{k_\mathrm{B}T}]^2\exp[\frac{\Delta E(\mu_0H_{\parallel})}{k_\mathrm{B}T}]}{\{\exp[\frac{\Delta E(\mu_0H_{\parallel})}{k_\mathrm{B}T}]+1\}^2},
\label{eqS6}
\end{equation}
where $\Delta E(\mu_0H_{\parallel})$ = $\mu_\mathrm{B}g_{\eta}\mu_0H_{\parallel}$  is the energy gap between two Zeeman levels of free spins under the magnetic field of $H_{\parallel}$ applied along the $c$ axis. At 2 T, the calculated $C_f$/$T$ displays an obvious peak at $T$ = 0.9 K [see the green line in Fig.~\ref{fig_free}(b)], while the total measured one ($C_\mathrm{p}$/$T$) shows no sign of peaks below 1.5 K. Moreover, at 0.36 K the Schottky specific heat $C_f$/$T$ = 0.067 JK$^{-2}$/mol-Cu calculated at 0.4 T is even larger than the total ones, 0.038 and 0.043 JK$^{-2}$/mol-Cu measured at 0 and 2 T, respectively [Fig.~\ref{fig_free}(b)]. In the model widely used in other frustrated magnets, e.g., ZnCu$_3$(OH)$_6$Cl$_2$~\cite{PhysRevLett.100.157205} and [NH$_4$]$_2$[C$_7$H$_{14}$N][V$_7$O$_6$F$_{18}$]~\cite{PhysRevLett.110.207208}, both the lattice and ``intrinsic" magnetic specific heat is believed to be independent of applied magnetic field up to $\sim$ 9 T, and thus their contributions are expected to be completely canceled out in the subtracted data, $\Delta C_\mathrm{p}$/$T$ = [$C_\mathrm{p}(\mathrm{0~T})-C_\mathrm{p}(\mu_0H_{\parallel})$]/$T$, as shown in Fig.~\ref{fig_free}(c). However, the Schottky model [$C_f(\mathrm{0~T})-C_f(\mu_0H_{\parallel})$]/$T$, see Eq.~(\ref{eqS6}) for $C_f$] with $f_{\parallel}$ = 0.80\% clearly fails to capture the subtracted data of YCOB, with a rather large residual $R_p$ = 32.3 $\gg$ 1 [see Eq.~(\ref{eq0}) for the definition of $R_p$]. Following Refs.~\cite{PhysRevLett.100.157205,PhysRevLett.110.207208}, we also treat both $f_{\parallel}$ and $\Delta E(\mu_0H_{\parallel})$ as adjustable parameters, and obtain $f_{\parallel}$ = 0.04, $\Delta E$(0 T) = 7.4 K, $\Delta E$(2 T) = 7.7 K, $\Delta E$(6 T) = 9.5 K, and $\Delta E$(9 T) = 11.4 K with the least $R_p$ = 17.4 $\gg$ 1. In sharp contrast to ZnCu$_3$(OH)$_6$Cl$_2$~\cite{PhysRevLett.100.157205}, [NH$_4$]$_2$[C$_7$H$_{14}$N][V$_7$O$_6$F$_{18}$]~\cite{PhysRevLett.110.207208}, and ZnCu$_3$(OH)$_6$SO$_4$~\cite{li2014Gapless}, etc., the fitted $\Delta E(\mu_0H_{\parallel})$ in YCOB clearly deviates far from the expected Zeeman energy gap of free spins.

The above inconsistency unambiguously demonstrates that the concentration of orphan spins is much lower than 0.8\%. The slight polarization and weak upturn observed in the low-$T$ (below 0.2$J_1$) magnetization and susceptibilities, respectively, should be governed by the ``intrinsic" magnetic properties of the KHA YCOB. Based on the DFT + $U$ calculations discussed in Appendix~\ref{sec2}, the site mixing between OH and Br causes two alternate exchanges along 70\% of the hexagons on the kagome lattice, and leads to the random model no.2, which resolves the above inconsistency naturally [see main text for the reproduction of the weak upturn and slight polarization, and see Fig.~\ref{fig_free}(c) for the simulations of the specific heat differences above $\sim$ 7 K $\sim$ 0.1$J_1$].

Even when the Brillouin function for 0.8\% of the free spins is subtracted from the raw susceptibilities [see Fig.~\ref{fig_free}(d)], we find that the final refinement results keep almost unchanged~\footnote{The main changes are the increases of the fitted $J_1$ and least $R_p$ by $\sim$ 2 K and $\sim$ 0.6, respectively, after the subtraction.}, as the difference is really trivial above $T_{\mathrm{min}}$ = 6 K (see Fig.~\ref{figs7}). Therefore, we use the experimental raw susceptibilities to refine the exchange Hamiltonian below.

\subsection{Finite-temperature Lanczos diagonalization (FLD) method}
\label{sec3p1}

\begin{figure}
\begin{center}
\includegraphics[width=8.6cm,angle=0]{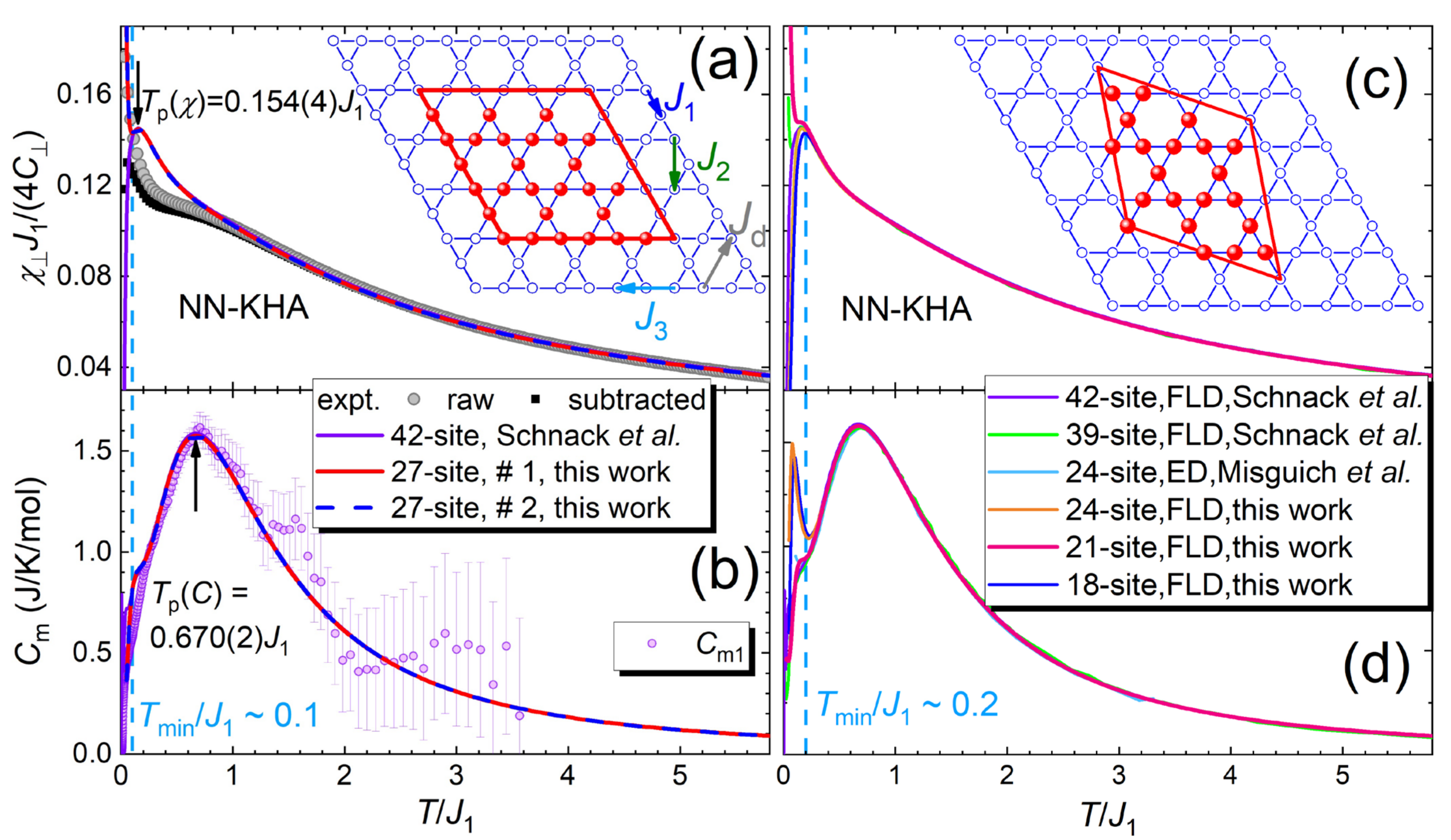}
\caption{Calculated magnetic susceptibility (a) and specific heat (b) of the 27-site NN-KHA cluster as function of normalized temperature ($T/J_1$) at zero field, with the reported 42-site FLD results~\cite{PhysRevB.98.094423} for comparison. The size effect is negligible at $T$ $\geq$ 0.1$J_1$. The inset of (a) shows the 27-site KHA cluster with the Heisenberg exchange interactions (colored arrows).  Calculated magnetic susceptibility (c) and specific heat (d) of the 24-, 21-, and 18-site NN-KHA clusters as function of normalized temperature at zero field, with the reported 24-site ED~\cite{Misguich2007}, 39- and 42-site FLD~\cite{PhysRevB.98.094423} results for comparison. The inset of (c) displays the 21-site KHA cluster. The size effect of susceptibility is negligible at $T$ $\geq$ 0.2$J_1$, whereas the specific heat shows size effect below $\sim$ 0.3$J_1$.}
\label{figs6}
\end{center}
\end{figure}
\begin{figure}
\begin{center}
\includegraphics[width=7cm,angle=0]{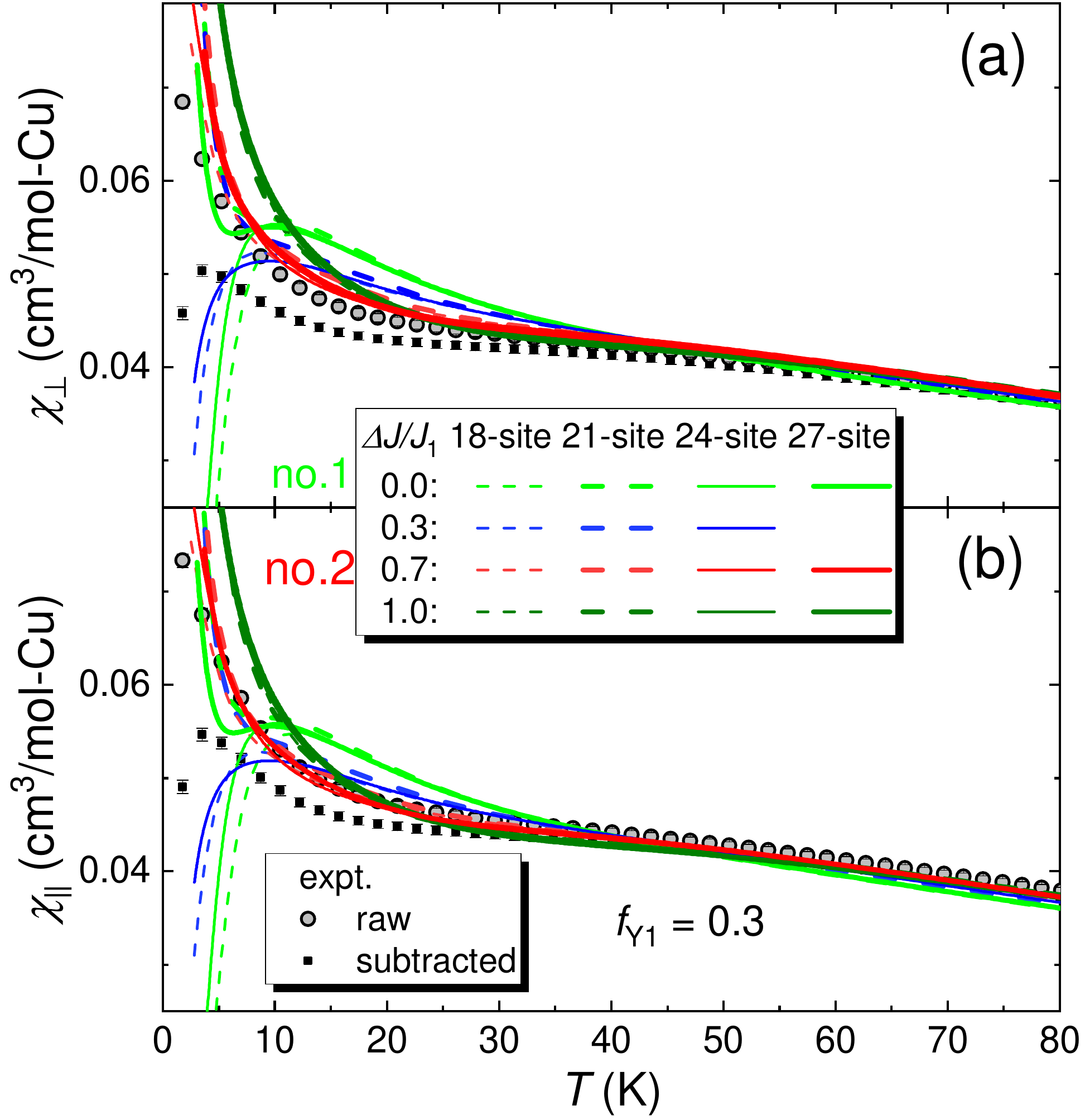}
\caption{(a, b) Least-$R_p$ fits to the experimental raw susceptibilities with different fixed fluctuations of exchange couplings, $\Delta J$/$J_1$, at $f_{\mathrm{Y}1}$ = 0.3 experimentally determined by XRD, above 6 K. The FLD calculations on various size random KHA clusters are conducted, and the finite-size effect is almost negligible at $\Delta J$/$J_1$ $\geq$ 0.7 (see a and b). For clarity, we present here only the low-$T$ data, please see main text for the full temperature range.}
\label{figs10}
\end{center}
\end{figure}
\begin{figure}
\begin{center}
\includegraphics[width=8.6cm,angle=0]{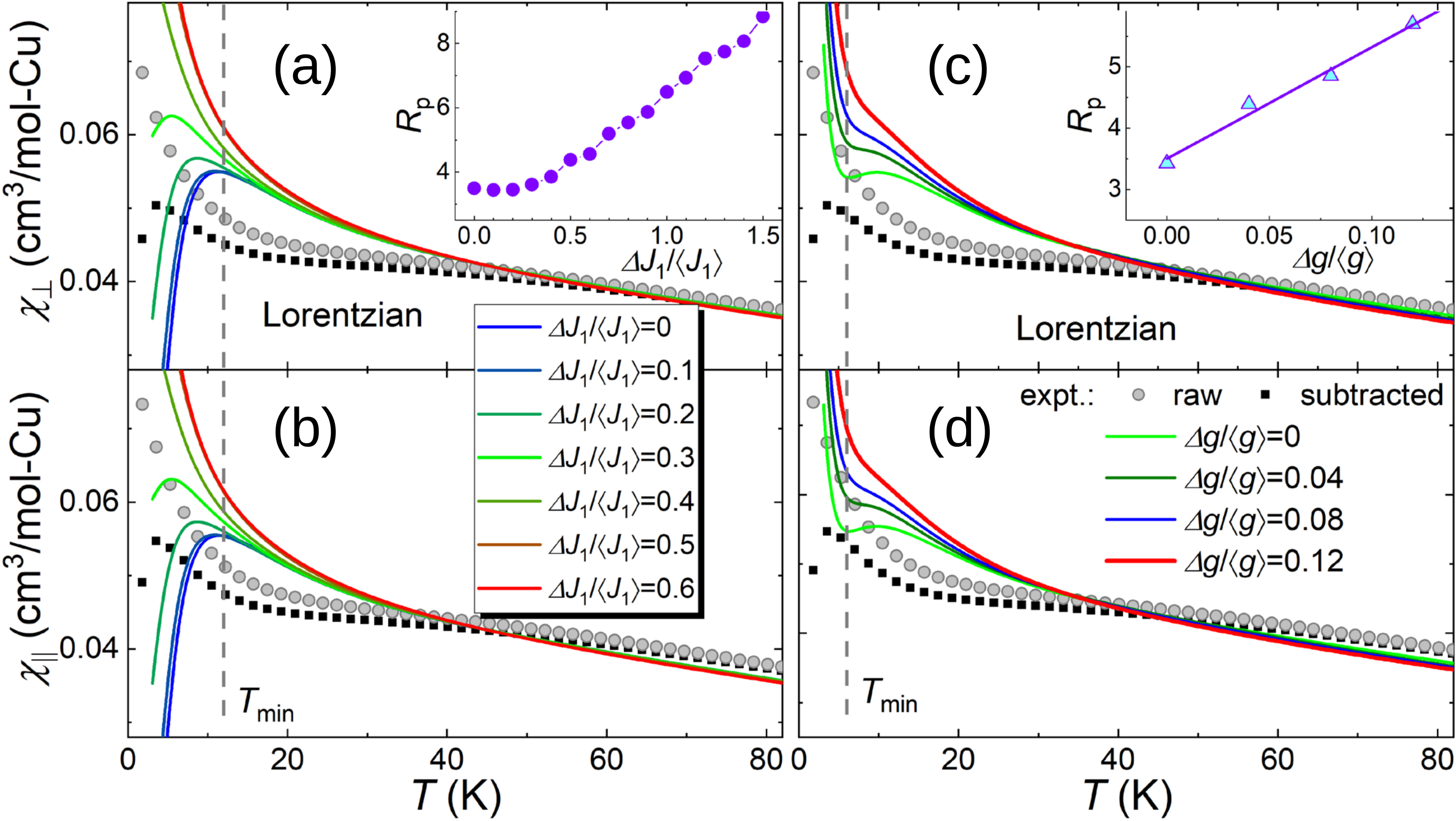}
\caption{(a, b) Calculated susceptibilities of the 18-site NN-KHA clusters with the Lorentzian distribution of the NN Heisenberg interactions. (c, d) Calculated susceptibilities of the 27-site NN-KHA clusters with the Lorentzian distribution of the $g$ factors. The experimental raw data of YCOB measured perpendicular and parallel to the $c$ axis are shown by gray circles, the data subtracted by the Brillouin functions for $\sim$ 0.8\% of free spins are displayed by black squares, and the dashed lines mark the lower valid bounds of the 18- and 27-site FLD simulations $T_{\mathrm{min}}$ = 0.2$J_1$ and 0.1$J_1$, respectively, in the ideal KHA case. The insets of (a, c) display the randomness strength dependence of the deviation $R_p$ (fit to data above $T_{\mathrm{min}}$).}
\label{figs7}
\end{center}
\end{figure}

We carried out FLD calculations for the thermodynamic properties on the 18-, 21-, 24-, and 27-site kagome clusters with periodic boundary conditions (PBC)~\cite{PhysRevB.49.5065,Liu2021Frustrated} [insets of Fig.~\ref{figs6}(a) and ~\ref{figs6}(c)]. 40 Lanczos steps and 10 different randomly chosen states were used in all our FLD calculations. The finite-size effect of susceptibility turns out to be negligible above $T_{\mathrm{min}}$ = 0.1$J_1$ $\sim$ 7 K and 0.2$J_1$ $\sim$ 12 K for the 27- and 21-(18-, 24-) site FLD calculations, respectively, with the error less than $\sim$ 1\% compared to the reported 42-site results~\cite{PhysRevB.98.094423} (see Fig.~\ref{figs6}). In our 27-site FLD calculations, $S^z\equiv\sum_{j}S_j^z$ symmetry of the system was used, and the exact diagonalization calculations were conduced in the subspaces of $|S^z|$ $\geq$ 23/2~\cite{PhysRevB.98.094423}. No obvious difference between two independent 27-site FLD calculations is found [see Fig.~\ref{figs6}(a) and \ref{figs6}(b)]. In the ideal $S$ = $\frac{1}{2}$ NN-KHA case, broad peaks of our calculated magnetic susceptibility and specific heat appear at $T_\mathbf{p}$($\chi$) = 0.154(4)$J_1$ and $T_\mathbf{p}$($C$) = 0.670(2)$J_1$, respectively, which are well consistent with the previously reported results~\cite{PhysRevB.98.094423}.

The ideal $S$ = $\frac{1}{2}$ NN-KHA model can't capture precisely the magnetic susceptibilities measured on YCOB below $T$ $\sim$ 0.6$J_1$ $\sim$ 40 K [see Figs.~\ref{figs6}(a)]. To better understand the thermodynamic observations, we consider a more complicated spin Hamiltonian with bond randomness. In this section, we first fixed the NN exchange coupling at $J_1$ = 1, and calculated the thermodynamic quantities with varying strength of the perturbations. Thereupon, we simultaneously fit the magnetic susceptibilities of YCOB measured both parallel and perpendicular to the $c$ axis above $T_{\mathrm{min}}$ through fine-tuning $J_1$ and $g$ by minimizing the following loss function [see Eq.~(\ref{eq0})].

\subsection{Bond randomness.}
\label{sec3p2}

Based on the crystal structure determined by XRD (see Table~\ref{table1}), as well as our DFT + $U$ calculations (see Appendix~\ref{sec2}), we constructed the model no.2 with randomly distributed hexagons of alternate bonds (see main text), and performed the FLD calculations on the 18-, 21-, 24-, and 27-site clusters with PBC. The calculated thermodynamic data are evaluated over 80, 80, 40, and 36 independent samples for the 18-, 21-, 24-, and 27-site FLD calculations, respectively, that have led to fully converged results of the random KHA ($X_i^{\mathrm{cal}}$) (see Fig.~\ref{figs10}). The fluctuation of the bonds ($\Delta J$/$J_1$) gradually induces local moments, which account for the growth of the upturn in susceptibility and the decrease of the finite-size effect of the FLD calculations. At $\Delta J$/$J_1$ $\geq$ 0.7, the finite-size effect almost disappears and our FLD simulations get very convincing even at low temperatures ($T\sim$ 0.1$J_1$). Both the weak upturn below $\sim$ 0.5$J_1$ and broad hump at $\sim$ $J_1$ seen in susceptibilities can be excellently reproduced by this model (see Fig.~\ref{figs10}).

If the local symmetries of the bond randomness aren't taken into account and conventional Lorentzian distributions of the exchange couplings and $g$ factors are introduced (the average values are $\langle J_1\rangle$ and $\langle g\rangle$, and the full widths at half maximum are $\Delta J_1$ and $\Delta g$)~\cite{li2015gapless,PhysRevLett.118.107202,Li2019YbMgGaO4,PhysRevX.10.011007,PhysRevB.97.184434,Li2021spin}, there's no way one can better fit the magnetic susceptibilities measured on YCOB, and the residual $R_p$ increases monotonically with the complete and continuous randomness (see Figs.~\ref{figs7}). Because the complete and continuous randomness fails to reproduce the characteristic hump seen in susceptibilities at $\sim$ $J_1$. The above analysis further validates our microscopic exchange model no.2 based on the crystal structure and DFT + $U$ calculations for YCOB.

\section{Magnetic specific heat and entropy.}
\label{sec4}

\begin{figure}
\begin{center}
\includegraphics[width=8.6cm,angle=0]{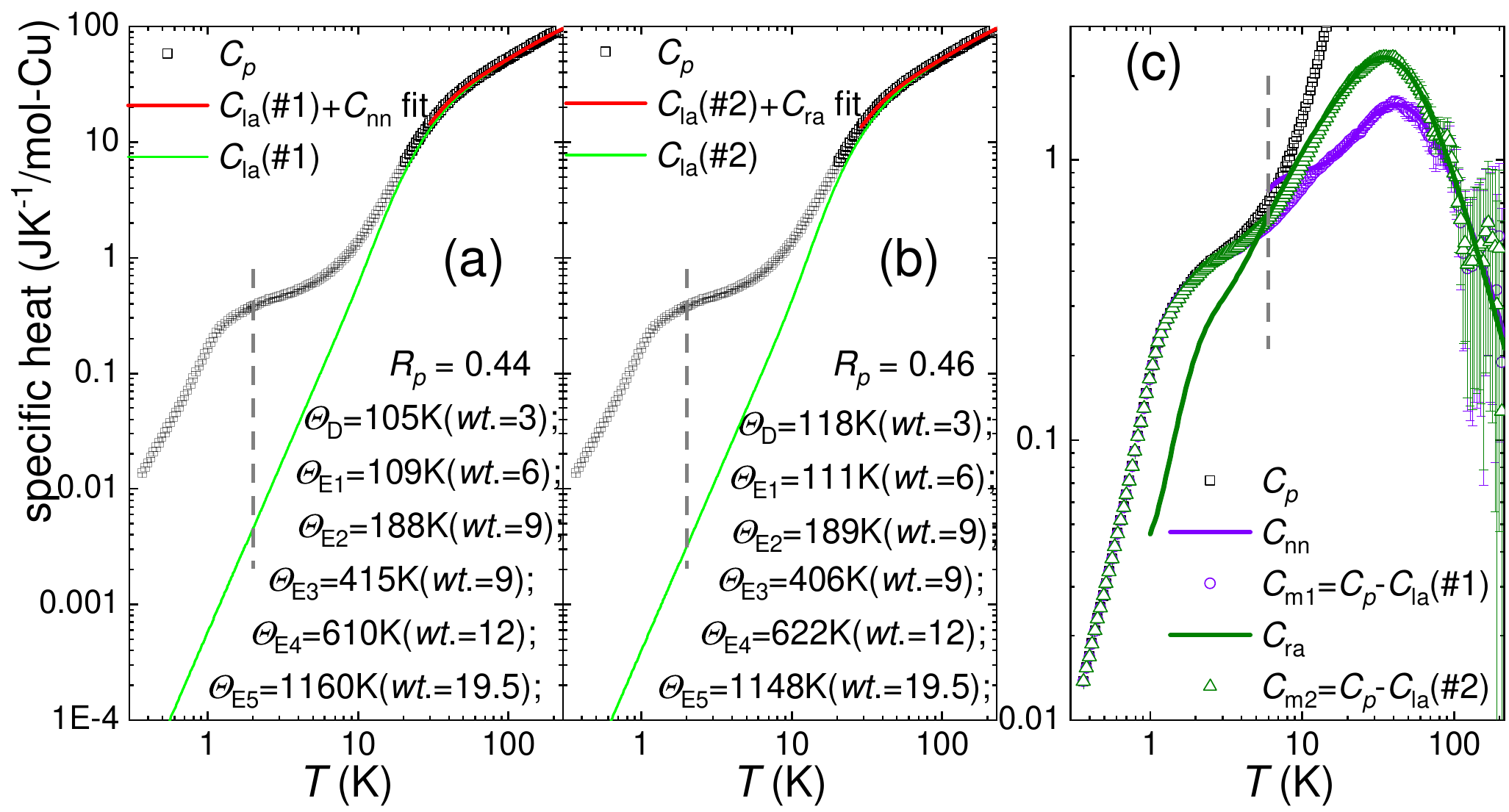}
\caption{(a, b) Temperature dependence of the total specific heat ($C_\mathrm{p}$, per mol Cu$^{2+}$) measured on the YCOB single crystal at 0 T. The red lines show the least-$R_p$ fit using the fixed model nos.1 (pure NN-KHA, see a) and 2 (with randomly distributed hexagons of alternate bonds, see b) and the green lines present the resulted lattice contributions ($C_{\mathrm{la}}$). (c) Magnetic specific heat of YCOB, with $C_\mathrm{p}$ for comparison. The violet and olive lines display the specific heat calculated by using models nos.1 ($C_{\mathrm{nn}}$) and 2 ($C_{\mathrm{ra}}$), respectively.}
\label{figs11}
\end{center}
\end{figure}
\begin{figure}
\begin{center}
\includegraphics[width=8.6cm,angle=0]{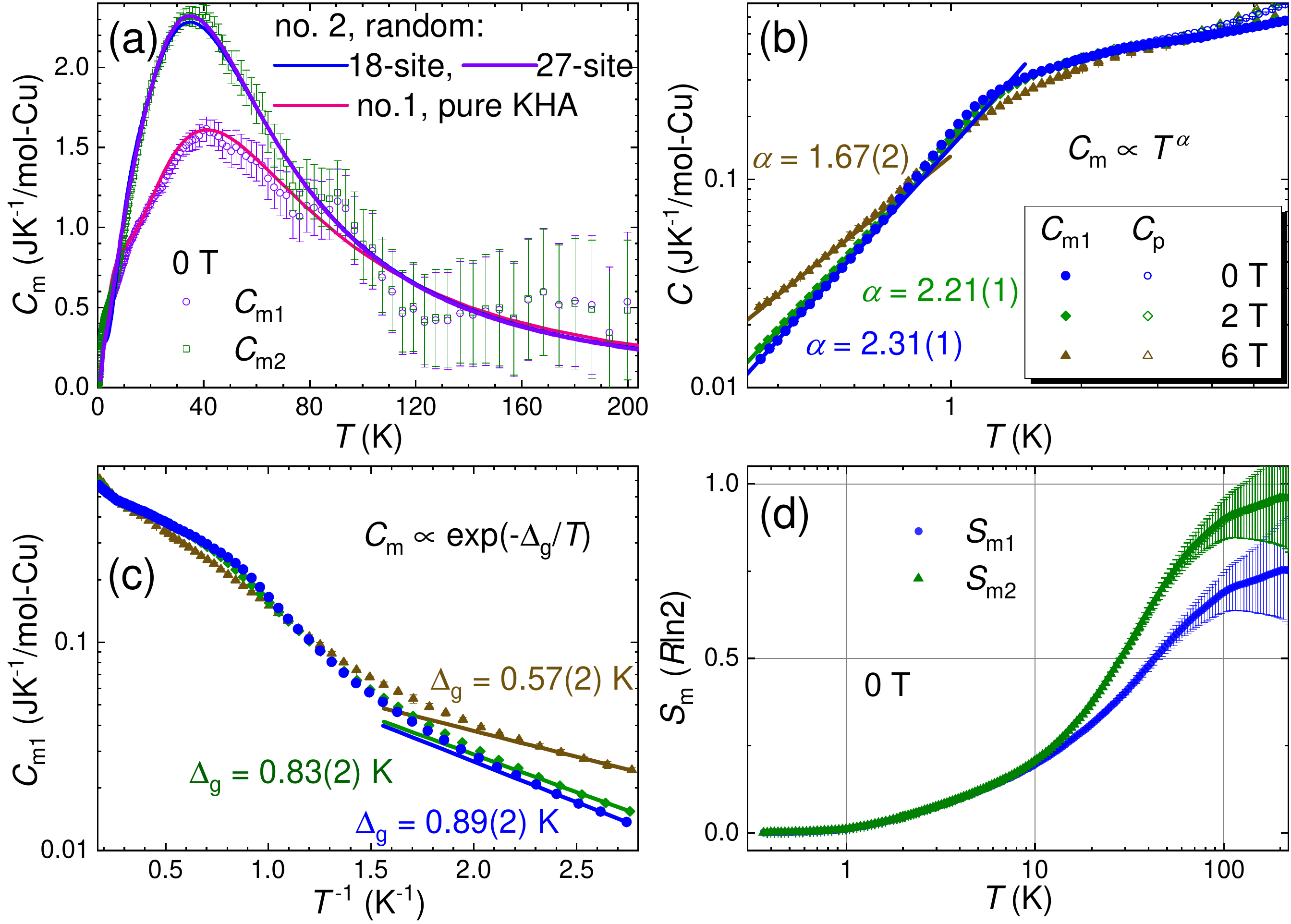}
\caption{(a) Magnetic specific heat of YCOB ($C_{\mathrm{m1}}$ and $C_{\mathrm{m2}}$) obtained by subtracting the refined lattice contributions from the total one (see Fig.~\ref{figs11}). The colored lines represent the specific heat calculated by using the model nos. 1 and 2 (see main text). (b) The magnetic specific heat of YCOB ($C_{\mathrm{m1}}$) with the colored lines showing the power-law fits. The raw $C_\mathrm{p}$ data are shown for comparison. (c) The corresponding $C_{\mathrm{m1}}$ vs $T^{-1}$ plot with the colored lines showing the exponential fits below 0.4 K. (d) Magnetic entropy of YCOB, $S_{\mathrm{m1}}$ and $S_{\mathrm{m2}}$, obtained from integrating $C_{\mathrm{m1}}/T$ and $C_{\mathrm{m2}}/T$, respectively.}
\label{figs12}
\end{center}
\end{figure}

Above $\sim$ 10 K, extracting precise magnetic specific heat from the total measured one is an extremely challenging task, in the absence of nonmagnetic reference compounds. Recently, a relatively convincing method had been reported in the frustrated  magnets YCu$_3$(OH)$_6$Cl$_3$~\cite{PhysRevLett.125.027203} and $\alpha$-CrOOH(D)~\cite{Liu2021Frustrated}. In this method, one first fits the total measured specific heat by using a combination of a typical Debye-Einstein lattice model and the fixed FLD result (the spin Hamiltonian determined by magnetization/susceptibility measurements is fixed) well above $T_{\mathrm{min}}$. After subtracting the refined Debye-Einstein lattice contribution, one obtains the magnetic specific heat in the full temperature range.

In the case of YCOB, we use the 27-site FLD specific heat results calculated from the pure NN-KHA ($C_{\mathrm{nn}}$, $J_1$ = 61.9 K, model no.1) and random  Hamiltonians($C_{\mathrm{ra}}$, $J_1$ = 50 K and $\Delta J$ = 0.7$J_1$, model no.2), and choose to fit the total measured specific heat above 30 K through tuning the Debye and Einstein temperatures [see Figs.~\ref{figs11}(a) and \ref{figs11}(b)]. Each unit cell of YCOB has $\sim$ 19.5 atoms (see Table~\ref{table1}), and thus 3 acoustic and $\sim$ 55.5 optical vibration modes. Therefore, we use the similar Debye-Einstein function previously reported in Ref.~\cite{Liu2021Frustrated} for YCOB,
\begin{multline}
3C_{\mathrm{la}}=\frac{9RT^3}{\Theta_\mathrm{D}^3}\int_0^{\frac{\Theta_\mathrm{D}}{T}}\frac{\xi^4e^{\xi}}{(e^{\xi}-1)^2}\mathrm{d}\xi\\
+\frac{R}{T^2}\sum_{n=1}^{5}\frac{w_n\Theta_{\mathrm{E}n}^2e^{\frac{\Theta_{\mathrm{E}n}}{T}}}{(e^{\frac{\Theta_{\mathrm{E}n}}{T}}-1)^2},
\label{eqS7}
\end{multline}
where $\Theta_\mathrm{D}$ and $\Theta_{\mathrm{E}n}$ are fitting parameters. From fitting, we obtain two series of very similar Debye and Einstein temperatures [$\Theta_\mathrm{D}$ and $\Theta_{\mathrm{E}n}$ ($n$ = 1, 2, 3, 4, 5), listed in Figs.~\ref{figs11}(a) and \ref{figs11}(b)] by fixing $w_1$, $w_2$, $w_3$, $w_4$, $w_5$ = 6, 9, 9, 12, 19.5, respectively. We further calculate (extrapolate) the Debye-Einstein lattice specific heat using Eq.~(\ref{eqS7}) in the full temperature range [see Figs.~\ref{figs11}(a) and \ref{figs11}(b)], and get the magnetic specific heat by removing the lattice contributions~\cite{PhysRevLett.125.027203}. As shown in Fig.~\ref{figs11}(c), the resulted magnetic specific heat, $C_{\mathrm{m1}}$ and $C_{\mathrm{m2}}$, indeed show little dependence upon the starting KHA model below $\sim$ 6 K. To be completely unbiased, we never use the specific heat data above 2 K to refine the spin Hamiltonian throughout this work.

\begin{figure}
\begin{center}
\includegraphics[width=8.6cm,angle=0]{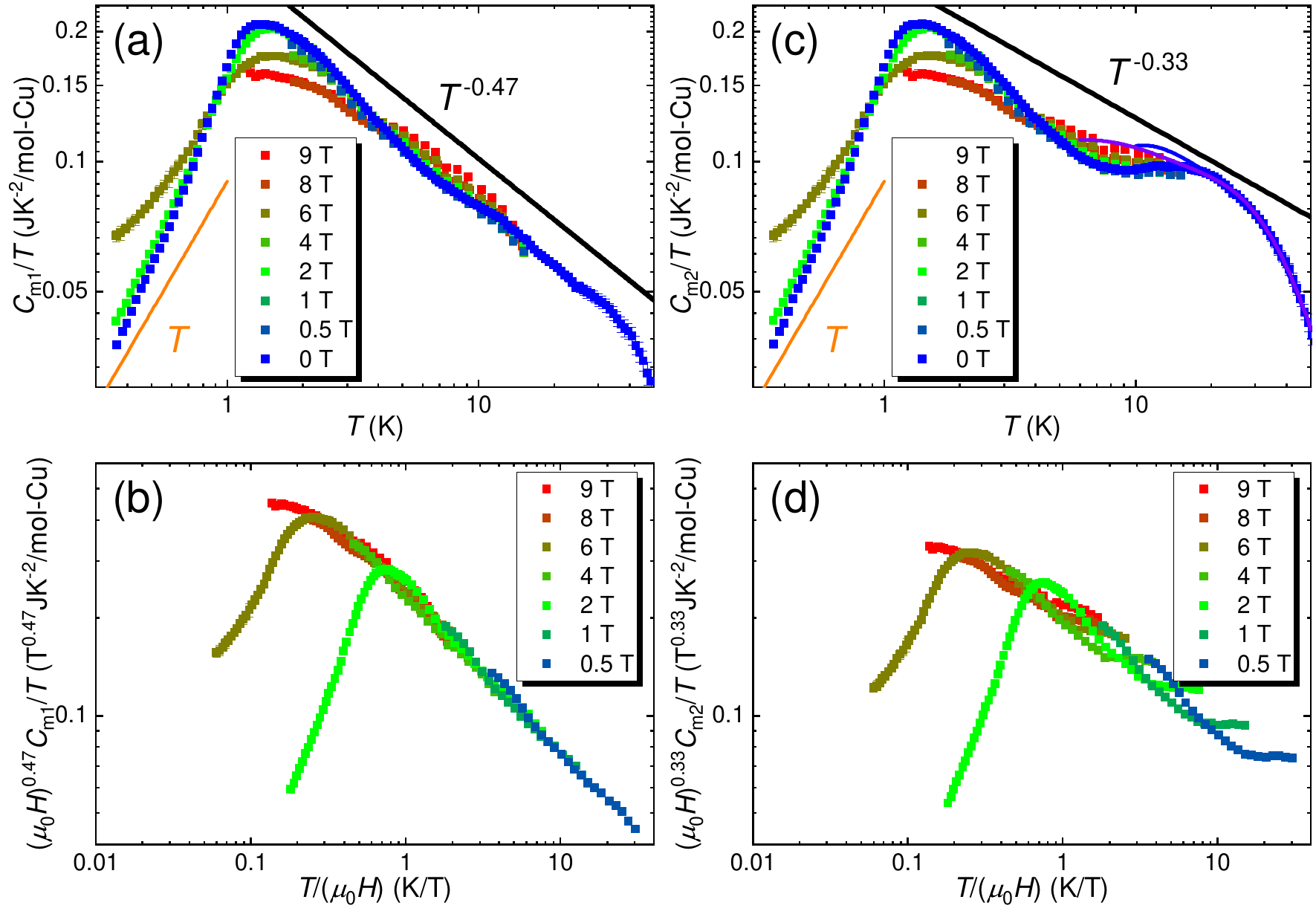}
\caption{(a, c) Temperature dependence of magnetic specific heat $C_{\mathrm{m1}}$ and $C_{\mathrm{m2}}$ under various magnetic fields applied along the $c$ axis. Above $\sim$ 1.2 K, the zero-field data roughly show powder-law behaviors, $C_{\mathrm{m1}}$/$T$ $\sim$ $T^{-0.47}$ (a) and $C_{\mathrm{m2}}$/$T$ $\sim$ $T^{-0.33}$ (c), respectively. The orange lines mark the quadratic behavior $C_{\mathrm{m}}$/$T$ $\propto$ $T$. (b, d) Data collapse in YCOB observed above $\sim$ 1.2 K. }
\label{figs13}
\end{center}
\end{figure}

There is no difference between $C_{\mathrm{m1}}$ and $C_{\mathrm{m2}}$ below $\sim$ 6 K, and thus we only show $C_{\mathrm{m1}}$ in Figs.~\ref{figs12}(b) and \ref{figs12}(c). Below 2 K, the lattice specific heat is completely negligible, which is more than two orders of magnitude smaller than the total $C_\mathrm{p}$ measured on YCOB [see Figs.~\ref{figs11}(a) and \ref{figs11}(b)]. The (magnetic) specific heat of YCOB shows a power-law behavior, $C_{\mathrm{m}}$ $\propto$ $T^{\alpha}$ at $T$ $<$ 1 K [see Fig.~\ref{figs12}(b)]. Alternatively, one could fit the low-temperature specific heat assuming an exponential behavior, $C_{\mathrm{m}}$ $\propto$ $\exp(-\Delta_\mathrm{g}/T)$, but such a fit extends only up to 0.4 K with a very small gap $\Delta_\mathrm{g}$ $\leq$ 0.9 K = 0.015$J_1$ [see Fig.~\ref{figs12}(c)]. Below $\sim$ 1 K, the temperature dependence of magnetic entropy flattens out at $S_{\mathrm{m}}$ $\leq$ 1.3\%$R\ln2$ [see Fig.~\ref{figs12}(d)], suggesting that we are indeed accessing the GS properties. The final increase of the magnetic entropy of using model no.2 from 0.36 to 220 K is more close to the expected value of $R\ln2$ based on the third law of thermodynamics than those of using other models, possibly further confirming the validation of the random model no.2 in YCOB.

\section{Scaling plot of specific heat.}
\label{sec5}

Recently, Kimchi et al. have developed a theory for the scaling collapse based on an emergent random-singlet regime in frustrated disordered quantum spin systems~\cite{Kimchi2018Scaling}. This theory is applicable to lots of spin-$\frac{1}{2}$ quantum magnets including LiZn$_2$Mo$_3$O$_8$ and ZnCu$_3$(OH)$_6$Cl$_2$, where the low-$T$ specific heat behaves as $C$($H,T$)/$T$ $\sim$ $H^{-\gamma}F_q$($T$/$H$). Here, $F_q$ is a general scaling function~\cite{Kimchi2018Scaling}.

Above $\sim$ 0.02$J_1$ $\sim$ 1.2 K, the temperature and field dependence of the magnetic heat capacity of YCOB are highly similar to those of LiZn$_2$Mo$_3$O$_8$ (see Fig.~\ref{figs13})~\cite{Kimchi2018Scaling}, and indeed show a scaling collapse with a familiar $\gamma$ $\sim$ 0.33--0.47 (see Fig.~\ref{figs13}) owing to the bond randomness. However, below $\sim$ 0.02$J_1$ $\sim$ 1.2 K the zero-field magnetic specific heat of YCOB $C_{\mathrm{m}}$/$T$ increases with the temperature, $C_{\mathrm{m}}$/$T$ $\sim$ $T^{1.31(1)}$ [see Fig.~\ref{figs12}(b)], giving rise to a negative $\gamma$ = $-$1.31(1). However, the theory for the scaling collapse is based on the non-universal exponent 0 $\leq$ $\gamma$ $\leq$ 1, and thus obviously isn't applicable to YCOB below the crossover temperature of $\sim$ 0.02$J_1$ $\sim$ 1.2 K [see Figs.~\ref{figs13}(b) and \ref{figs13}(d)].

\section{Mean-field Ansatz of various spin-liquid states.}
\label{sec6}

In this section, we try to understand the specific heat behavior of YCOB below $\sim$ 2 K, where the lattice contribution is completely negligible and the raw measured data directly reflect the GS properties of the spin system. Since no conventional magnetic transition is found in YCOB, we start with the effective mean-field models (i.e. mean-field Ansatz) of various QSLs, including the uniform resonating-valence-bond (RVB), U(1) Dirac, and $\mathbb{Z}_2$ states on the kagome lattice.

\begin{figure*}
\begin{center}
\includegraphics[width=16cm,angle=0]{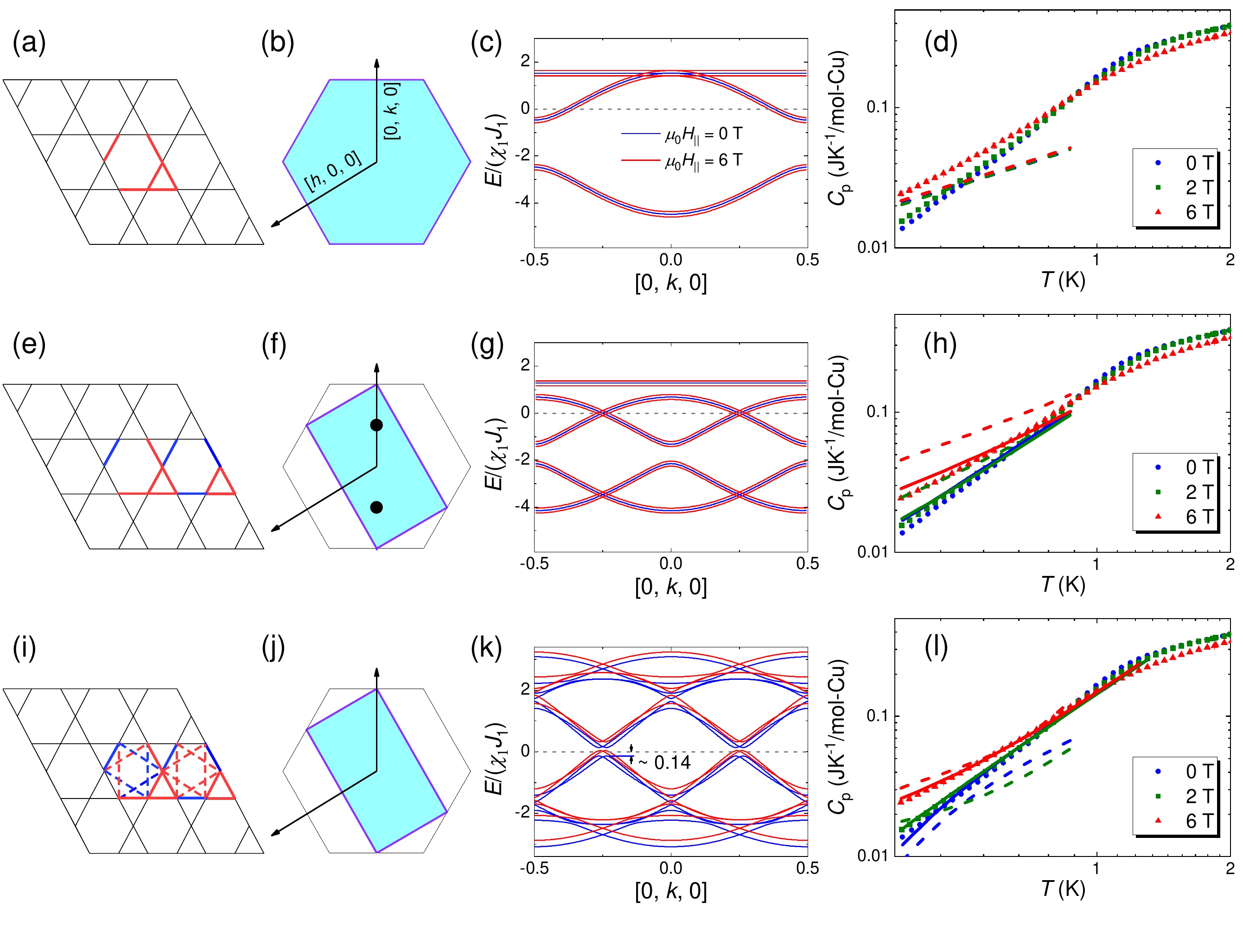}
\caption{(a, e, i) Kagome lattice with showing the sign structures of hopping terms in a unit cell. Red and blue solid lines present the first-neighbor $\nu_{jl}$ = 1 and $-$1, whereas the red and blue dashed lines (i) present the second-neighbor $\nu_{jl'}$ = 1 and $-$1, respectively. (b, f, j) The first Brillouin zone for the uniform RVB, U(1) Dirac, and $\mathbb{Z}_2[0,\pi]\beta$ QSL states. The outer hexagons show the first Brillouin zone for the kagome lattice with the arrows showing the [$h$, 0, 0] and [0, $k$, 0] directions. The black circles mark the positions of the Dirac nodes in (f). (c, g, k) The band structures of the uniform RVB, U(1) Dirac, and $\mathbb{Z}_2[0,\pi]\beta$ QSL states along the [0, $k$, 0] direction at $h$ = 0. The calculated band structures in 0 and 6 T are shown by the blue and red lines, respectively. In the cases of (c) uniform RVB and (g) U(1) Dirac QSLs, the spin-up and -down bands of spinons are degenerate in zero field (the flat band marked in blue is fourfold degenerate in g), and are split in an applied magnetic field (the flat bands marked in red are still doubly degenerate in g). In the case of $\mathbb{Z}_2[0,\pi]\beta$ state (k), the applied field shifts the energy levels up by $\mu_0H_{\parallel}\mu_{\mathrm{eff}}$. (d, h, l) Low-$T$ specific heat of YCOB measured in magnetic fields applied along the $c$ axis, with colored lines showing the calculations using the mean-field Ansatz of uniform RVB, U(1) Dirac, and $\mathbb{Z}_2[0,\pi]\beta$ states. The solid lines display the fits with the adjustable effective magnetic moment $\mu_{\mathrm{eff}}$ $\sim$ 0.5$\mu_{\mathrm{B}}$, whereas the dashed lines show the results with the fixed $\mu_{\mathrm{eff}}$ = $\mu_{\mathrm{B}}$.}
\label{figs14}
\end{center}
\end{figure*}

\subsection{Uniform resonating-valence-bond state.}
\label{sec6p0}

The simplest mean-field Hamiltonian is the so-called uniform RVB state and is given by~\cite{PhysRevLett.98.117205},
\begin{multline}
\mathcal{H}_{\mathrm{uniform}}=-J_1[\lambda_3\sum_{j\sigma} f_{j\sigma}^{\dag}f_{j\sigma}+\chi_1\sum_{\langle jl\rangle,\sigma}(f_{j\sigma}^{\dag}f_{l\sigma}+\mathrm{H.c.})]\\
-\mu_0H_{\parallel}\mu_\mathrm{eff}\sum_{j,\sigma}\sigma f_{j\sigma}^{\dag}f_{j\sigma},
\label{eqS6:1}
\end{multline}
where $\sigma$ = $\uparrow$ ($+$1) and $\downarrow$ ($-$1) represents spin-up and -down fermionic spinons, respectively, $\mu_\mathrm{eff}$ is the effective spin-$\frac{1}{2}$ moments, and $\chi_1$ is the first-neighbor hopping term [see Fig.~\ref{figs14}(a)]. The chemical potential $\lambda_3$ in both the uniform RVB and U(1) Dirac scenarios are self-consistently determined by the constraint, $\sum_{j}(\sum_{\sigma}\langle f_{j\sigma}^{\dag}f_{j\sigma}\rangle-1)$ = 0, at each temperature and in each applied magnetic field~\cite{PhysRevLett.98.117205}, and $J_1$ = 60 K is fixed in this section. It is worth mentioning that the above constraint isn't strict enough to impose the single-occupancy constraint $\sum_{\sigma} f_{j\sigma}^{\dag}f_{j\sigma}$ = 1~\cite{RevModPhys.89.025003}. Therefore, in the framework of the mean-field theory many unphysical states can contribute to extra density of states and thus specific heat, which may account for the reduced specific heat coefficient $\gamma_2$ and effect magnetic moment $\mu_\mathrm{eff}$ observed in YCOB (see below).

Below $\sim$ 1 K, the zero-field specific heat of YCOB exhibits a nearly quadratic temperature dependence, $C_{\mathrm{\mathrm{p}}}$ = 0.142(1)$T^{2.31(1)}$ (JK$^{-1}$/mol-Cu), which contradicts with the uniform RVB state. A linear temperature dependence of specific heat is expected in the uniform RVB state, because of the large Fermi surface [see Fig.~\ref{figs14}(c)]. The applied magnetic field up to $\sim$ 6 T only slightly changes the density of states around Fermi surface, and thus the calculated field dependence of specific heat is extremely weak [see Fig.~\ref{figs14}(d)]. In contrast, the specific heat measured on YCOB shows a clear magnetic field dependence [see Fig.~\ref{figs14}(d)], and thus unambiguously precludes the uniform RVB scenario.

\subsection{U(1) Dirac spin liquid.}
\label{sec6p1}

Similarly, the mean-field Ansatz of the U(1) SL state is given by~\cite{PhysRevLett.98.117205},
\begin{multline}
\mathcal{H}_{\mathrm{U(1)}}=-J_1[\lambda_3\sum_{j,\sigma}f_{j\sigma}^{\dag}f_{j\sigma}+\chi_1\sum_{\langle jl\rangle,\sigma}\nu_{jl}(f_{j\sigma}^{\dag}f_{l\sigma}+\mathrm{H.c.})]\\
-\mu_0H_{\parallel}\mu_\mathrm{eff}\sum_{j,\sigma}\sigma f_{j\sigma}^{\dag}f_{j\sigma},
\label{eqS6:2}
\end{multline}
where $\nu_{jl}$ $\equiv$ $\pm$1 presents the sign structure of hopping terms [see Fig.~\ref{figs14}(e)]. Our calculated band structure is identical to that reported in Ref.~\cite{PhysRevLett.98.117205}, and the temperature and field dependence of low-$T$ specific heat in both $\mu_0H_{\parallel}\mu_\mathrm{eff}$ $\ll$ $k_\mathrm{B}T$ $\ll$ $\chi_1J_1$ and $k_\mathrm{B}T$ $\ll$ $\mu_0H_{\parallel}\mu_\mathrm{eff}$ $\ll$ $\chi_1J_1$ limits are reproduced.

At 0 T, the low-$T$ specific heat of U(1) SL state behaves as $C_\mathrm{p}$ = $\frac{72\zeta(3)\pi k_\mathrm{B}^3A}{(2\pi\hbar v_\mathrm{F})}T^2$ (JK$^{-1}$/mol-Cu)~\cite{PhysRevLett.98.117205},
where $v_\mathrm{F}$ = $\frac{a\chi_1J_1}{\sqrt{2}\hbar}$ is the Fermi velocity and $A$ = $\frac{\sqrt{3}N_\mathrm{A}a^2}{6}$ = 7.72$\times$10$^4$ m$^2$/mol-Cu is the area of the two-dimensional system. From fitting the specific heat using the quadratic-temperature function below $\sim$ 1 K, we obtain $C_\mathrm{p}\sim\gamma_2T^2$ with $\gamma_2$ = 0.11(1) JK$^{-3}$/mol-Cu at 0 T, and thus $v_\mathrm{F}$ = 1.07(5)$\times$10$^3$ m/s, in YCOB. The resulted $\chi_1$ = 0.29(1) is roughly consistent with the mean-field value of $\chi_1(\mathrm{MF})$ = 0.221 found by Hastings~\cite{PhysRevB.63.014413}, and thus supports the U(1) Dirac scenario in YCOB. The mean-field theory gives a slightly overestimated specific heat coefficient $\gamma_2(\mathrm{MF})$ = $\frac{6\sqrt{3}\zeta(3)k_\mathrm{B}^3N_\mathrm{A}}{\pi J_1^2[\chi_1(\mathrm{MF})]^2}$ = 0.188 JK$^{-3}$/mol-Cu, possibly owing to the unphysical states (see above).

When a magnetic field is applied, the band structure tends to form Fermi pockets around the Dirac nodes in the reciprocal space [see Fig.~\ref{figs14}(g)]. Therefore, both low-energy density of states and thus low-$T$ specific heat increase with the strength of the magnetic field. The spinons with the free electron moment $\mu_{\mathrm{eff}}$ = $\mu_{\mathrm{B}}$ will cause a stronger dependence of specific heat upon the magnetic field, and obviously can't explain the observations, resulting in a large $R_p$ = 127 [see Fig.~\ref{figs14}(h)]. A remedy is to set $\mu_{\mathrm{eff}}$ an adjustable parameter. From fitting all the low-$T$ ($<$ 1 K) specific heat data of YCOB simultaneously, we obtain $\chi_1$ = 0.27 and a significantly reduced moment $\mu_{\mathrm{eff}}$ = 0.43$\mu_{\mathrm{B}}$ with the least $R_p$ = 20.4 [see Fig.~\ref{figs14}(h)].

\subsection{$\mathbb{Z}_2[0,\pi]\beta$ spin liquid.}
\label{sec6p2}

Finally, we check the $\mathbb{Z}_2$ scenario for YCOB, and use the mean-field Hamiltonian previously reported in Ref.~\cite{PhysRevB.83.224413},
\begin{multline}
\mathcal{H}_{\mathbb{Z}_2}=-J_1[\sum_j(\lambda_3\sum_\sigma f_{j\sigma}^{\dag}f_{j\sigma}+\lambda_1f_{j\uparrow}^{\dag}f_{j\downarrow}^{\dag}+\mathrm{H.c.})\\
+\chi_1\sum_{\langle jl\rangle,\sigma}\nu_{jl}(f_{j\sigma}^{\dag}f_{l\sigma}+\mathrm{H.c.}\\
+\sum_{\langle\langle jl'\rangle\rangle}\nu_{jl}(\chi_2\sum_\sigma f_{j\sigma}^{\dag}f_{l'\sigma}+\Delta_2\sum_\sigma\sigma f_{j\sigma}^{\dag}f_{l'\bar{\sigma}}^{\dag}+\mathrm{H.c.})]\\
-\mu_0H_{\parallel}\mu_\mathrm{eff}\sum_{j,\sigma}\sigma f_{j\sigma}^{\dag}f_{j\sigma},
\label{eqS6:3}
\end{multline}
where $\lambda_1$ and $\lambda_3$ are single-ion chemical potentials self-consistently determined by $\sum_{j}\langle f_{j\uparrow}^{\dag}f_{j\downarrow}^{\dag}\rangle$ = $\sum_{j}\langle f_{j\uparrow}f_{j\downarrow}\rangle$ = 0 and $\sum_{j}(\sum_{\sigma}\langle f_{j\sigma}^{\dag}f_{j\sigma}\rangle-1)$ = 0, and the mean-field Ansatz up to the second neighbors in a unit cell is shown in Fig.~\ref{figs14}(i). By setting the mean-field parameters $\chi_2$ = $\Delta_2$ = 0, we naturally go back to the U(1) Dirac scenario [see Eq.~(\ref{eqS6:2})], and reproduce the band structure~\footnote{It is worth mentioning that there is a flip for the spin-down band since we use the Nambu representation in the $\mathbb{Z}_2[0,\pi]\beta$ scenario.} and observable quantities (e.g., specific heat) of the U(1) Dirac QSL. When both $\chi_2$ and $\Delta_2$ are small, the $\mathbb{Z}_2[0,\pi]\beta$ state is in the neighborhood of the U(1) Dirac SL state, and has a small gap due to the singlet-pairing term $\Delta_2$ $\neq$ 0 around the wave vectors of the previous U(1) Dirac nodes.

Similarly, from fitting the specific heat data below $\sim$ 1 K we obtain $\chi_1$ = 0.176, $\chi_2$ = 0.043, $\Delta_2$ = 0.013, and $\mu_{\mathrm{eff}}$ = 0.48$\mu_{\mathrm{B}}$ with the least $R_p$ = 7.9 [see Fig.~\ref{figs14}(l)]. The $\Delta_2$ term opens a small spin (triplet) gap $\sim$ 0.14$\chi_1J_1$ $\sim$ 0.025$J_1$ in the band structure [see Fig.~\ref{figs14}(k)]. When $\mu_{\mathrm{eff}}$ = $\mu_{\mathrm{B}}$ is fixed, we get $\chi_1$ = 0.160, $\chi_2$ = 0.11, and $\Delta_2$ = 0.014, with $R_p$ = 32.4 [see Fig.~\ref{figs14}(l)].

\bigbreak

\bibliography{cu_kag}

\end{document}